\documentclass[12pt]{iopart}

\expandafter\let\csname equation*\endcsname\relax

\expandafter\let\csname endequation*\endcsname\relax

\usepackage{newtxtext,newtxmath}

\usepackage{graphicx}
\usepackage{xcolor}
\makeatletter
\newcommand{\colorcaption}[2][]{%
  \begingroup%
  \renewcommand{\@caption@fignum@sep}{ (Color online). }%
  \caption[#1]{#2}%
  \endgroup%
}
\makeatother

\usepackage{upgreek}
\newcommand{\hi}[1]{\textcolor{black}{#1}}

\usepackage{hyperref}

\begin{document}

    %%-------------------------------------------------------------------------------------------

	%%\title{Superconducting triplet pairing in Al/Al$ _\mathbf{2} $O$ _\mathbf{3} $(/EuS)/Ni/Ga junctions}
	\title{\hi{Signatures of superconducting triplet pairing in Ni--Ga-bilayer junctions}}

	\author{Andreas Costa}%
 	\address{Institute for Theoretical Physics, University of Regensburg, 93040 Regensburg, Germany}
 	\ead{andreas.costa@physik.uni-regensburg.de}

	\author{Madison Sutula}%
 	\address{Francis Bitter Magnet Laboratory and Plasma Science and Fusion Center, Massachusetts Institute of Technology, MA 02139, USA}
 	\address{Department of Materials Science and Engineering, Massachusetts Institute of Technology, MA 02139, USA}

 	\author{Valeria Lauter}%
 	\address{Neutron Scattering Division, Neutron Sciences Directorate, Oak Ridge National Laboratory, Oak Ridge, TN 37831, USA}
 	\ead{lauterv@ornl.gov}

 	\author{Jia Song}%
 	\address{Francis Bitter Magnet Laboratory and Plasma Science and Fusion Center, Massachusetts Institute of Technology, MA 02139, USA}

 	\author{Jaroslav Fabian}%
 	\address{Institute for Theoretical Physics, University of Regensburg, 93040 Regensburg, Germany}
 	\ead{jaroslav.fabian@physik.uni-regensburg.de}

 	\author{Jagadeesh S. Moodera}%
 	\address{Francis Bitter Magnet Laboratory and Plasma Science and Fusion Center, Massachusetts Institute of Technology, MA 02139, USA}
 	\address{Department of Physics, Massachusetts Institute of Technology, MA 02139, USA}
 	\ead{moodera@mit.edu}

 	%\vspace{10pt}
    %\begin{indented}
    %\item[] \hi{February 2022}
    %\end{indented}

 	%%\date{\today}
    
    %%-------------------------------------------------------------------------------------------
    
    \begin{abstract}
    	Ni--Ga~bilayers are a versatile platform for exploring the competition between strongly antagonistic ferromagnetic and superconducting phases. 
    	We characterize the impact of this competition on the transport~properties of highly-ballistic Al/Al$ _2 $O$ _3 $(/EuS)/Ni--Ga tunnel junctions from both experimental and theoretical points of view. 
    	While the conductance spectra of junctions comprising Ni ($ 3 \, \mathrm{nm} $)--Ga ($ 60 \, \mathrm{nm} $) bilayers can be well understood within the framework of earlier results, which associate the emerging main conductance maxima with the junction~films' superconducting gaps, \hi{thinner $ \mathrm{Ni} $ ($ 1.6 \, \mathrm{nm} $)--$ \mathrm{Ga} $ ($ 30 \, \mathrm{nm} $) bilayers entail completely different physics, and give rise to novel large-bias~(when compared to the superconducting gap of the thin Al~film as a reference) conductance-peak subseries that we term \emph{conductance~shoulders}. These conductance~shoulders might attract considerable attention also in similar magnetic superconducting bilayer junctions, as we predict them to offer an experimentally well-accessible transport signature of superconducting triplet pairings that are induced around the interface of the Ni--Ga~bilayer. We further substantiate this claim performing complementary polarized~neutron~reflectometry measurements on the bilayers, from which we deduce (1)~a \emph{nonuniform magnetization~structure} in~Ga in a several nanometer-thick area around the Ni--Ga~boundary and can simultaneously (2)~satisfactorily fit the obtained data only considering the \emph{paramagnetic Meissner response} scenario. While the latter provides independent experimental evidence of induced triplet superconductivity inside the Ni--Ga~bilayer, the former might serve as the first experimental hint of its potential microscopic physical origin. Finally, we introduce a simple phenomenological toy model to confirm also from the theoretical standpoint that superconducting triplet pairings around the Ni--Ga~interface can indeed lead to the experimentally observed conductance~shoulders, which convinces that our claims are robust and physically justified. Arranging our work in a broader context, we expect that Ni--Ga-bilayer junctions could have a strong potential for future superconducting-spintronics applications whenever an efficient engineering of triplet-pairing superconductivity is required. }
    \end{abstract}
    
    %%-------------------------------------------------------------------------------------------
    
    %%\maketitle
    
    %%-------------------------------------------------------------------------------------------

    \vspace{2pc}
    \noindent{\it Keywords}: superconducting triplet~pairing, polarized~neutron reflectometry, paramagnetic Meissner~response 
    %\submitto{\NJP}
    %\maketitle
    
    \section{Introduction   \label{Sec_Intro}}

    Superconducting magnetic junctions form elementary building~blocks for superconducting~spintronics~\cite{Fabian2004,Fabian2007,Eschrig2011,Linder2015,Ohnishi2020}, with potential applications in quantum~computing~\cite{Ioffe1999,Mooij1999,Blatter2001,Ustinov2003,Yamashita2005,Feofanov2010,Khabipov2010,Devoret2013}. 
    Early conductance measurements on ferromagnet/superconductor 
    point~contacts~\cite{Soulen1998,Soulen1999} demonstrated that Andreev~reflection can be used to quantify the ferromagnet's spin~polarization~\cite{DeJong1995}. 
    Nowadays, more complex structures, such 
    as magnetic Josephson-junction~geometries~\cite{Golubov2004}, in which Yu--Shiba--Rusinov~states~\cite{Yu1965,Shiba1968,Rusinov1968,Rusinov1968alt} can strongly influence the supercurrent~\cite{Costa2018,Kochan2020} and even induce current-reversing $0$-$\pi$~transitions~\cite{Bulaevskii1977a,Bulaevskii1977b,Ryazanov2001}, are being exploited. 
    A wealth of unique physical phenomena and transport~anomalies has been predicted to emerge in such junctions, covering the potential formation of Majorana~states~\cite{Nilsson2008,Duckheim2011,Lee2012a,Nadj-Perge2014,Dumitrescu2015,Pawlak2016,Ruby2017,Livanas2019,Manna2020}, significantly magnified current~magnetoanisotropies~\cite{Hoegl2015,Hoegl2015a,Jacobsen2016,Costa2017,Martinez2020}, as well as the efficient generation and detection of spin-polarized triplet Cooper-pair~currents~\cite{Keizer2006,Eschrig2011}.

    Particularly appealing materials for superconducting spintronics are Ni--Ga~(Ni--Bi)~bilayers~\cite{Moodera1990,LeClair2005}, as strong proximity~effects turn the intrinsically weakly ferromagnetic Ni~film superconducting. 
    Coexistence of two nominally antagonistic ferromagnetic and superconducting phases in the Ni~film can strongly modify transport~properties, such as differential~conductance. 
    Most remarkable is the possibility of generating spin-triplet states, as previous studies~\cite{Belzig1996,Kadigrobov2001,Bergeret2001,Bergeret2005,Keizer2006,Yokoyama2007a,Linder2009,Khaire2010,Robinson2010,Anwar2010,Yokoyama2011a,Bergeret2012,Bergeret2013,Bergeret2014,Alidoust2015a,DiBernardo2015,Arjoranta2016,Espedal2016,Pal2017,Bergeret2020} drew the conclusion that ferromagnetic exchange can induce odd-frequency superconductivity as a signature of triplet~pairing.

    The two main factors that cause triplet~pairing in proximitized $ s $-wave superconductors are inhomogeneously magnetized domains and spin-orbit~coupling~effects.
    While triplet currents originating from nonuniform magnetizations have been successfully implemented in various systems~\cite{Bergeret2001b,Bergeret2001,Bergeret2001a,Eschrig2003,Houzet2007,Eschrig2008,Grein2009,Khaire2010,Robinson2010a,Robinson2010,Banerjee2014,Diesch2018}---e.g., in Nb/Py/Co/Py/Nb~junctions through tilting the thin permalloy~(Py)~spin-mixers' magnetizations~\cite{Banerjee2014}---, \hi{generating them through spin-orbit~coupling could, in certain cases~\cite{Bergeret2013,Bergeret2014}, become more challenging and require specific magnetization~configurations~(relative to the spin-orbit~field) to induce sizable enough triplet~pairings~\cite{Satchell2018,Satchell2019,Eskilt2019,Bujnowski2019}}.

    In this paper, we \hi{experimentally} investigate the tunneling~conductance~(dominated by quasiparticles) of high-quality superconducting magnetic Al/Al$ _2 $O$ _3 $(/EuS)/Ni--Ga~junctions that host either thicker Ni~($ 3 \, \mathrm{nm} $)--Ga~($ 60 \, \mathrm{nm} $) or thinner $ \mathrm{Ni} $~($ 1.6 \, \mathrm{nm} $)--$ \mathrm{Ga} $~($ 30 \, \mathrm{nm} $)~bilayers. 
    In the latter case, we observe a series of unexpected \emph{conductance~shoulders} forming at rather large bias~voltages, \hi{when compared to the superconducting gap of the thin Al~film as a reference, and predict that these conductance~shoulders serve as a possible transport fingerprint of superconducting triplet pairings being induced around the Ni--Ga~interface. 
    To give independent experimental evidence of this claim, and gain first insights into the potentially underlying physics, we conduct complementary polarized neutron reflectometry studies on specific Ni--Ga~bilayers, which eventually allow us to visualize the magnetization around the peculiar Ni--Ga~boundary. 
    Thereby detecting a nonuniformly magnetized area around the Ni--Ga~interface, as well as the \emph{paramagnetic Meissner~response} in Ga, provides the key experimental evidence that even weak intrinsic ferromagnetism in Ni can induce a superconducting triplet state~\cite{Machida1978,Yokoyama2011a,Alidoust2014,Asano2014,DiBernardo2015a} near the interface of superconducting Ni--Ga~bilayers. 
    We further substantiate our findings by means of a a simple phenomenological theoretical toy model that demonstrates that considering superconducting triplet~pairings near the Ni--Ga~interface is indeed sufficient to qualitatively recover the experimentally observed conductance~shoulders. 
    Moreover, we briefly comment on samples with thinner Ni--Ga~bilayers that contain an additional, strongly spin-polarized, EuS~barrier. The latter is expected to substantially enhance the ferromagnetic exchange~interaction inside the junctions, and notably modify their transport~characteristics. }

    We have structured the paper as follows. 
    In~Sec.~\ref{Sec_Exp}, we briefly summarize our state-of-the-art techniques to grow the high-quality superconducting magnetic tunnel~junctions, present and discuss the results of our tunneling-conductance~measurements carried out on selected samples---\hi{paying special attention to the yet puzzling novel large-bias conductance~shoulders}---, \hi{and finally analyze the results of our polarized~neutron~reflectometry measurements.}  
    Section~\ref{Sec_Theory} reports on our theoretical efforts to develop a simple, \hi{\emph{purely phenomenological}}, description that relates the large-bias conductance~shoulders to induced superconducting triplet~pairings near the Ni--Ga-bilayer~interface. 
    Finally, we briefly conclude our main findings in~Sec.~\ref{Sec_Conclusions}. 
    \hi{Our results might provide an essential contribution to establish Ni--Ga~bilayers as promising platforms to engineer spin-polarized triplet~supercurrents in future works. }

    \section{Experimental study: Conductance features and polarized neutron reflectometry  \label{Sec_Exp}} 
    
    \subsection{Sample growth}

        %% Fig. JUNCTION SCHEMATICS %%
        \begin{figure}[tb]
            \centering
            \includegraphics[width=0.45\textwidth]{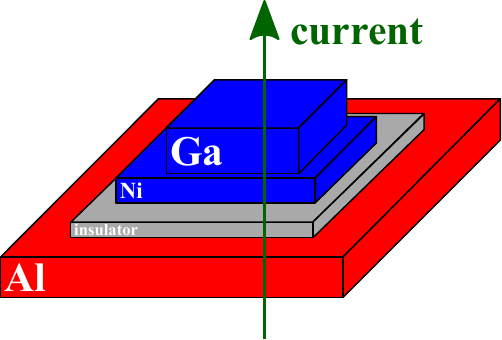}
            \caption{Sketch of the studied Al/insu\-la\-tor~[Al$ _2 $O$ _3 $~(/EuS)]/Ni--Ga~junctions. 
                The Al and Ga~electrodes are intrinsically superconducting, while proximity~effects additionally turn the intrinsically weakly ferromagnetic Ni~film likewise superconducting. 
                By applying a bias~voltage~$ V $ between the electrodes and measuring the corresponding tunneling~current~(indicated by a dark-green arrow), we probe the tunneling-conductance~characteristics of the samples. 
                    \label{FigStacking}}
        \end{figure}

        All investigated Al/Al$ _2 $O$ _3 $(/EuS)/Ni--Ga~junctions---schematically illustrated in~Fig.~\ref{FigStacking}---with cross-section areas of~$ 150 \, \upmu \mathrm{m} \times 150 \, \upmu \mathrm{m} $ were prepared by means of thermal evaporation inside an ultra-high~vacuum~(UHV)~system with a base~pressure of~$ 2 \times 10^{-8} \, \mathrm{mb} $ using in-situ shadow-masking~techniques. 
        During the growth~process, thin layers of Al, Ni, Ga, and (partly) EuS were evaporated on clean glass~substrates kept at temperatures of about~$ 80 \, \mathrm{K} $. 
        Ultrathin Al$ _2 $O$ _3 $~tunneling~barriers, separating adjacent Al and (EuS/)Ni--Ga~films, were created in~situ either by exposing Al to a controlled oxygen~plasma~(producing Al$ _2 $O$ _3 $~barriers about $ 1 \, \mathrm{nm} $~thick) or deposited from an $ \mathrm{Al}_2 \mathrm{O}_3 $~source using electron-beam~evaporation~(to obtain thinner $ 0.4 \, \mathrm{nm} $ Al$ _2 $O$ _3 $~barriers). 
        Before taking the junctions out of the UHV~chamber, they were protected by $ 12 \, \mathrm{nm} $-thick Al$ _2 $O$ _3 $~layers.

        In one run, we could prepare several junctions differing only in the thicknesses of individual Al, EuS, Ni, or Ga~layers, while keeping all other growth~parameters the same. 
        To measure the tunneling~conductance, and thus study the ramifications of the superconducting magnetic Ni--Ga~bilayers on transport, we attached the junctions to a probe with electrical leads and immersed the system into a pumped liquid-helium~bath~(either $ ^4 \mathrm{He} $ or $ ^3 \mathrm{He} $) to reach a temperature of about~$ 1 \, \mathrm{K} $ or $ 0.6 \, \mathrm{K} $. 
        Both Al and Ni--Ga thin films turned superconducting, with critical~temperatures strongly dependent on their thickness. 
        \hi{All our samples were based on $ 4 \, \mathrm{nm} $-thick Al~films~(serving as the left electrode; please note that studying the tunneling~conductance requires two distinct superconducting electrodes, which is in stark contrast to STM~studies~\cite{Diesch2018} that could probe the Ni--Ga~bilayers alone), which have themselves already been intensively investigated in Al/EuS/Al~junctions earlier~\cite{Rouco2019a} and demonstrated to remain superconducting below a critical temperature of about~$ T^\mathrm{crit.}_\mathrm{Al} \approx 2.5 \, \mathrm{K} $; the superconducting coherence length of such thin Al~films was estimated by Meservey and Tedrow~\cite{Meservey1994} to be~$ \xi_\mathrm{Al} \approx 50 \, \mathrm{nm} $. }
        \hi{Due to the strong competition between ferromagnetism and superconductivity, the critical temperature of the Ni--Ga~bilayers~(serving as the right electrode) is mostly determined by the thickness of the Ni~film, and usually drops down with increasing Ni~thickness~(stronger ferromagnetic exchange). At about $ 2\, \mathrm{nm} $ Ni, the critical temperature of the Ni--Ga~bilayer is roughly~$ T^\mathrm{crit.}_\text{Ni--Ga} \approx 4.2 \, \mathrm{K} $; typical superconducting coherence lengths of Ni~($ 0.4 \, \mathrm{nm} $)--Ga~($ 14 \, \mathrm{nm} $)~bilayers are of the order of~$ \xi_\text{Ni--Ga} \approx 15 \, \mathrm{nm} $~\cite{Moodera1990}. 
        Performing Meservey--Tedrow~spectroscopy~\cite{Tedrow1971,Tedrow1973,Meservey1994}, we further estimated that the weak intrinsic ferromagnetic exchange within thin Ni~films causes spin~polarizations of about~$ 1\% $, which noticeably increase above~$ 4 \, \mathrm{nm} $ Ni~thickness. }

        %% Tab. Samples %%
        \begin{table}[b]
                \renewcommand{\arraystretch}{1.2}
                \caption{Junction~composition of samples~A through D. 
                    In samples~A through C, the $ \mathrm{Al}_2 \mathrm{O}_3 $~barriers with a thickness of about~$ 1 \, \mathrm{nm} $, separating Al and Ni~films, were created by exposing Al to controlled oxygen~plasma. 
                    In sample~D, a $ 0.4 \, \mathrm{nm} $-thick $ \mathrm{Al}_2 \mathrm{O}_3 $~barrier was deposited from an $ \mathrm{Al}_2 \mathrm{O}_3 $~source using electron-beam~evaporation. %, and decouples the ferromagnetic exchange of EuS from the neighboring Al~electrode. 
                    }
				\label{TabSamples}
				\centering
				\begin{tabular}{l l}
				    \br
					\textbf{sample~A} & Al~($ 4 \, \mathrm{nm} $)/Al$ _2 $O$ _3 $~($ \sim 1 \, \mathrm{nm} $)/ \\
					& \hspace{45 pt} Ni~($ 3 \, \mathrm{nm} $)--Ga~($ 60 \, \mathrm{nm} $)/$ \mathrm{Al}_2 \mathrm{O}_3 $~($ 12 \, \mathrm{nm} $) \\[0.2 cm]
                    \textbf{sample~B} & $ \mathrm{Al} $~($ 4 \, \mathrm{nm} $)/$ \mathrm{Al}_2 \mathrm{O}_3 $~($ \sim 1 \, \mathrm{nm} $)/ \\
                    & \hspace{45 pt} $ \mathrm{Ni} $~($ 1.6 \, \mathrm{nm} $)--$ \mathrm{Ga} $~($ 30 \, \mathrm{nm} $)/$ \mathrm{Al}_2 \mathrm{O}_3 $~($ 12 \, \mathrm{nm} $) \\[0.2 cm]
                    \textbf{sample~C} & $ \mathrm{Al} $~($ 4 \, \mathrm{nm} $)/$ \mathrm{Al}_2 \mathrm{O}_3 $~($ \sim 1 \, \mathrm{nm} $)/ \\
                    & \hspace{45 pt} $ \mathrm{Ga} $~($ 30 \, \mathrm{nm} $)--$ \mathrm{Ni} $~($ 1.6 \, \mathrm{nm} $)/$ \mathrm{Al}_2 \mathrm{O}_3 $~($ 12 \, \mathrm{nm} $) \\[0.2 cm]
                    \textbf{sample~D} & $ \mathrm{Al} $~($ 4 \, \mathrm{nm} $)/$ \mathrm{Al}_2 \mathrm{O}_3 $~($ 0.4 \, \mathrm{nm} $)/$ \mathrm{EuS} $~($ 1.2 \, \mathrm{nm} $)/ \\
                    & \hspace{45 pt} $ \mathrm{Ni} $~($ 1.6 \, \mathrm{nm} $)--$ \mathrm{Ga} $~($ 30 \, \mathrm{nm}) $/$ \mathrm{Al}_2 \mathrm{O}_3 $~($ 12 \, \mathrm{nm} $) \\
                    \br
				\end{tabular}
	    \end{table}

    \subsection{Tunneling-conductance measurements}

        To demonstrate the most puzzling transport~features of Ni--Ga-bilayer junctions that we could detect in our series of experiments, we focus on four different samples~(labeled \emph{sample~A} through \emph{sample~D}; see~Tab.~\ref{TabSamples}) containing either a thicker Ni~($ 3 \, \mathrm{nm} $)--Ga~($ 60 \, \mathrm{nm} $) or a thinner $ \mathrm{Ni} $~($ 1.6 \, \mathrm{nm} $)--$ \mathrm{Ga} $~($ 30 \, \mathrm{nm} $)~bilayer, respectively. 
        All tunneling-conductance~data was obtained using standard lock-in~technique.

        %% Fig. Sample A %%
        \begin{figure}[tb]
            \centering
            \includegraphics[width=0.50\textwidth]{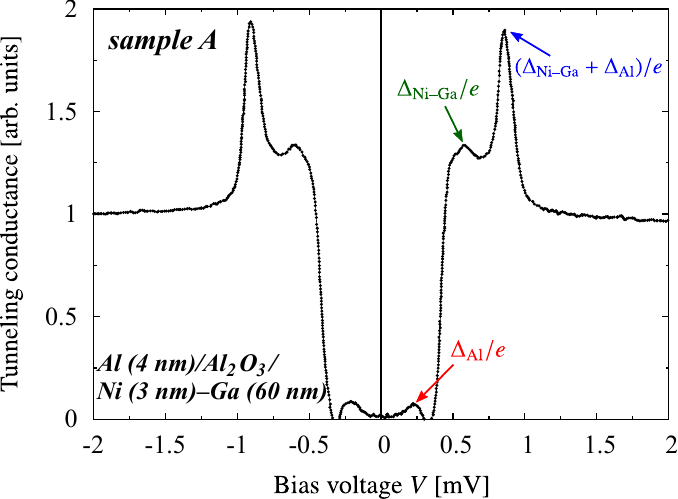}
            \caption{\hi{The \emph{measured} tunneling~conductance--bias~voltage~characteristics of the Al~($ 4 \, \mathrm{nm} $)/Al$ _2 $O$ _3 $~($ \sim 1 \, \mathrm{nm} $)/Ni~($ 3 \, \mathrm{nm} $)--Ga~($ 60 \, \mathrm{nm} $)/$ \mathrm{Al}_2 \mathrm{O}_3 $~($ 12 \, \mathrm{nm} $)~junction, with a thicker Ni~($ 3 \, \mathrm{nm} $)--Ga~($ 60 \, \mathrm{nm} $)~bilayer, at a temperature of about~$ 1.1 \, \mathrm{K} $. 
                The positions~(bias~voltages) of the \emph{main} conductance~peaks~(maxima) are determined by the superconducting gaps of the left Al~electrode, $ \Delta_\mathrm{Al} $, and the right Ni--Ga-bilayer~electrode, $ \Delta_\text{Ni--Ga} $, as indicated. }
                \label{FigCondThickBilayer}}
        \end{figure}

        \emph{Sample~A.} 
            First, we studied the Al~($ 4 \, \mathrm{nm} $)/Al$ _2 $O$ _3 $~($ \sim 1 \, \mathrm{nm} $)/Ni~($ 3 \, \mathrm{nm} $)--Ga~($ 60 \, \mathrm{nm} $)/ $ \mathrm{Al}_2 \mathrm{O}_3 $~($ 12 \, \mathrm{nm} $)~junction with a thicker Ni~($ 3 \, \mathrm{nm} $)--Ga~($ 60 \, \mathrm{nm} $)~bilayer and at a temperature of about~$ 1.1 \, \mathrm{K} $. 
            \hi{The results of our measurements, which are shown in~Fig.~\ref{FigCondThickBilayer}, reveal three distinct \emph{main}~(first-order) conductance~maxima. }

            \hi{The quasiparticle tunneling conductance of similar (Josephson-like) junction geometries, consisting of two superconducting electrodes that are separated by a thin nonsuperconducting link, has already been intensively investigated in numerous systems and by several authors before~(see, e.g., Refs.~\cite{Taylor1963,Adkins1963,Adkins1964,VanHuffelen1993,Kuhlmann1994,Zimmermann1995,Rowell1964,Rowell1968,Klapwijk1982,Octavio1983,Flensberg1988}). 
            Rowell and Feldman~\cite{Rowell1968,Octavio1983} developed thereby one of the perhaps most fundamental theoretical descriptions of these S/N/S'~junctions' tunneling~conductance, assuming two \emph{dissimilar} superconducting electrodes S and S' connected by a thin nonsuperconducting N~link. 
            Their Rowell--Feldman~approach predicts the emergence of \emph{main} conductance~peaks whenever the applied bias~voltage~$ V $ satisfies $ eV \approx \mp \Delta_\mathrm{S'} $, $ eV \approx \mp ( \Delta_\mathrm{S} + \Delta_\mathrm{S'} ) $, or $ eV \approx \mp \Delta_\mathrm{S} $, where~$ \Delta_\mathrm{S} $~($ \Delta_\mathrm{S'} $) denotes the superconducting~gap of S~(S') and $ e $ refers to the positive elementary charge. 
            At the microscopic level, Arnold~\cite{Arnold1985,Arnold1987} related the appearance of these conductance~peaks to multiple Andreev~reflections that occur at these particular bias~voltages. 
            }

        %% Fig. Samples B and C %%
        \begin{figure}[tb]
            \centering
            \includegraphics[width=0.975\textwidth]{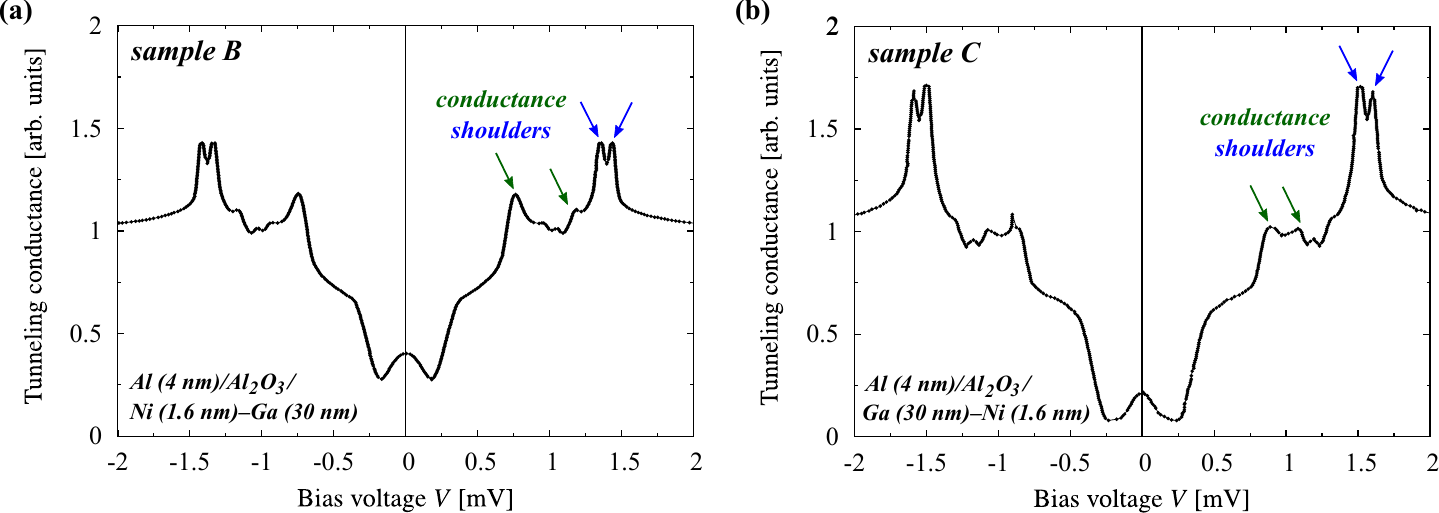}
            \caption{\hi{(a)~The \emph{measured} tunneling~conductance--bias~voltage~characteristics of the $ \mathrm{Al} $~($ 4 \, \mathrm{nm} $)/$ \mathrm{Al}_2 \mathrm{O}_3 $~($ \sim 1 \, \mathrm{nm} $)/$ \mathrm{Ni} $~($ 1.6 \, \mathrm{nm} $)--$ \mathrm{Ga} $~($ 30 \, \mathrm{nm} $)/$ \mathrm{Al}_2 \mathrm{O}_3 $~($ 12 \, \mathrm{nm} $)~junction, with a thinner $ \mathrm{Ni} $~($ 1.6 \, \mathrm{nm} $)--$ \mathrm{Ga} $~($ 30 \, \mathrm{nm} $)~bilayer, at a temperature of about~$ 1.1 \, \mathrm{K} $. 
                (b)~The same measurement as in~(a), but replacing the $ \mathrm{Ni} $~($ 1.6 \, \mathrm{nm} $)--$ \mathrm{Ga} $~($ 30 \, \mathrm{nm} $)~bilayer by an ``\emph{inverted}'' $ \mathrm{Ga} $~($ 30 \, \mathrm{nm} $)--$ \mathrm{Ni} $~($ 1.6 \, \mathrm{nm} $)~bilayer. 
                Note that the large-bias conductance~maxima, discussed in~Fig.~\ref{FigCondThickBilayer}, are divided into a series of conductance~shoulders indicated by green and blue arrows; each shoulder consists of (at least) two individually resolvable adjacent conductance~maxima. }
                %%In addition, a noticeable zero-bias conductance peak is formed~(purple arrows).
                    \label{FigCondThinBilayer}}
        \end{figure}

            \hi{Interestingly, we can directly adapt the predictions of the Rowell--Feldman~model to understand the conductance~features that we observed in sample~A~(see~Fig.~\ref{FigCondThickBilayer}). 
            We identify the thin Al~film on the left- and the Ni--Ga~bilayer on the right-hand side as the superconducting electrodes S and S' of our samples, while the thin Al$ _2 $O$ _3 $ tunneling~barrier serves as the nonsuperconducting link. 
            From the main conductance~peaks~(maxima) displayed in~Fig.~\ref{FigCondThickBilayer}, we can therefore estimate the superconducting gaps of Al, $ \Delta_\mathrm{Al} $, and Ni--Ga, $ \Delta_\text{Ni--Ga} $, such that the conductance~maxima arise at~$ eV \approx \mp \Delta_\mathrm{Al} $, $ eV \approx \mp \Delta_\text{Ni--Ga} $, and $ eV \approx \mp (\Delta_\mathrm{Al} + \Delta_\text{Ni--Ga}) $, accordingly. 
            At a temperature of about~$ 1.1 \, \mathrm{K} $, at which the conductance~measurements were performed, we finally obtain~$ \Delta_\mathrm{Al} \approx 0.25 \, \mathrm{meV} $---which is in good agreement with such thin Al~films' (zero-temperature) gap of about~$ 0.36 \, \mathrm{meV} $~\cite{Moodera1990}---as well as~$ \Delta_\text{Ni--Ga} \approx 0.6 \, \mathrm{meV} $; the latter might be compared to the gap of an earlier studied Ni~($ 2 \, \mathrm{nm} $)--Ga~($ 100 \, \mathrm{nm} $)~bilayer~\cite{Moodera1990}, which was estimated to be about~$ 0.57 \, \mathrm{meV} $ and is one of the rare references of Ni--Ga~bilayers that are available in the literature. 
            Since the Rowell--Feldmann~approach suffices to satisfactorily explain the conductance~features of sample~A, and extract physically reasonable values of the Al and Ni--Ga-bilayer electrodes' superconducting gaps, we might conclude already at this point that the ferromagnetic exchange inside Ni does not \emph{substantially} affect the physics of the thicker Ni~($ 3 \, \mathrm{nm} $)--Ga~($ 60 \, \mathrm{nm} $)~bilayer; recall that Rowell and Feldman did not account for ferromagnetic components. We will in fact see later on that the interface magnetization of thicker Ni--Ga~bilayers remains more uniform and, most likely due to their simultaneously weak spin~polarizations of just about~$ 1\% $, the Ni~films' ferromagnetism does consequently not yet raise novel physics in thicker Ni~($ 3 \, \mathrm{nm} $)--Ga~($ 60 \, \mathrm{nm} $)~bilayers. 
            }

        \emph{Samples~B and~C.} 
            The second and third samples that we investigated were composed of the $ \mathrm{Al} $~($ 4 \, \mathrm{nm} $)/$ \mathrm{Al}_2 \mathrm{O}_3 $~($ \sim 1 \, \mathrm{nm} $)/$ \mathrm{Ni} $~($ 1.6 \, \mathrm{nm} $)--$ \mathrm{Ga} $~($ 30 \, \mathrm{nm} $)/$ \mathrm{Al}_2 \mathrm{O}_3 $~($ 12 \, \mathrm{nm} $)~junction with a thinner $ \mathrm{Ni} $~($ 1.6 \, \mathrm{nm} $)--$ \mathrm{Ga} $~($ 30 \, \mathrm{nm} $)~bilayer and the $ \mathrm{Al} $~($ 4 \, \mathrm{nm} $)/$ \mathrm{Al}_2 \mathrm{O}_3 $~($ \sim 1 \, \mathrm{nm} $)/$ \mathrm{Ga} $~($ 30 \, \mathrm{nm} $)--$ \mathrm{Ni} $~($ 1.6 \, \mathrm{nm} $)/$ \mathrm{Al}_2 \mathrm{O}_3 $~($ 12 \, \mathrm{nm} $)~junction with an effectively ``\emph{inverted}'' $ \mathrm{Ga} $~($ 30 \, \mathrm{nm} $)--$ \mathrm{Ni} $~($ 1.6 \, \mathrm{nm} $)~bilayer, respectively. 
            \hi{The corresponding tunneling~conductances measured at a temperature of about~$ 1.1 \, \mathrm{K} $, which are shown in~Fig.~\ref{FigCondThinBilayer}, reflect much richer conductance~features than before~(in sample~A). }

            \hi{Specifically, each of the two large-bias~(taking the gap of the thin Al~film as the smallest energy reference in the system) conductance~maxima that we associated with~$ \Delta_\text{Ni--Ga} $ and~$ \Delta_\text{Ni--Ga} + \Delta_\mathrm{Al} $ in sample~A seems to split into a subseries of (at least) two distinct and individually resolvable conductance~maxima, which we will call \emph{conductance shoulders} hereinafter. 
            These conductance shoulders have not yet been explored in prior works; providing a sophisticated picture of their physical origin is therefore the main objective of our paper. 
            }

            \hi{As the conductance shoulders only arise in junctions with thinner Ni--Ga~bilayers, they are most likely intimately connected with novel physical phenomena that solely arise in thinner Ni--Ga~bilayers. 
            In our earlier theoretical work on the transport~characteristics of ferromagnet/superconductor/ferromagnet~junctions in the presence of interfacial spin-orbit~interactions~\cite{Costa2021}, we identified splittings of main conductance~peaks into a shoulder-like pattern as signatures of ``unconventional''~(i.e., spin-flip) Andreev~reflections at the interfaces. 
            These unconventional Andreev~reflections effectively induce superconducting triplet pairings in the junction. 
            From that point of view, the large-bias conductance shoulders occurring in junctions with thinner $ \mathrm{Ni} $~($ 1.6 \, \mathrm{nm} $)--$ \mathrm{Ga} $~($ 30 \, \mathrm{nm} $)~bilayers \emph{might} likewise provide transport~fingerprints of superconducting triplet pairings. 
            Nevertheless, before we can safely establish a connection between conductance shoulders and triplet superconductivity, we need to provide clear experimental evidence of the latter---which we will when analyzing the results of our polarized~neutron~reflectometry measurements. 
            }

        %% Fig. Sample D %%
        \begin{figure}[tb]
            \centering
            \includegraphics[width=0.50\textwidth]{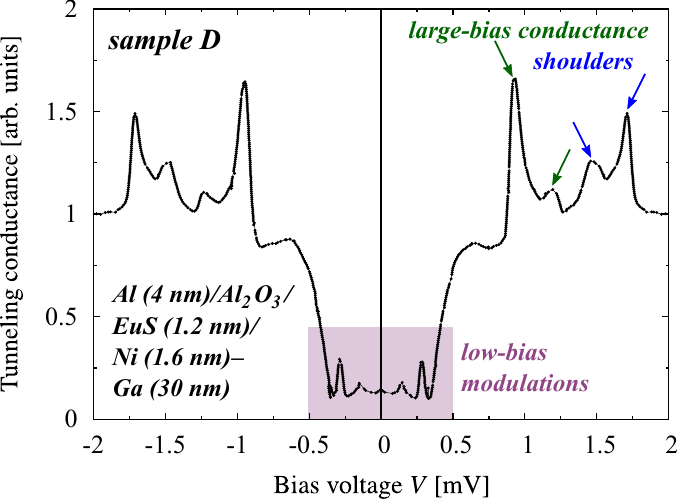}
            \caption{\hi{The \emph{measured} tunneling~conductance--bias~voltage~characteristics of the $ \mathrm{Al} $~($ 4 \, \mathrm{nm} $)/$ \mathrm{Al}_2 \mathrm{O}_3 $~($ 0.4 \, \mathrm{nm} $)/$ \mathrm{EuS} $~($ 1.2 \, \mathrm{nm} $)/$ \mathrm{Ni} $~($ 1.6 \, \mathrm{nm} $)--$ \mathrm{Ga} $~($ 30 \, \mathrm{nm}) $/$ \mathrm{Al}_2 \mathrm{O}_3 $~($ 12 \, \mathrm{nm} $)~junction, with a thinner $ \mathrm{Ni} $~($ 1.6 \, \mathrm{nm} $)--$ \mathrm{Ga} $~($ 30 \, \mathrm{nm} $)~bilayer, at a temperature of about~$ 1 \, \mathrm{K} $. Green and blue arrows indicate the large-bias conductance~shoulders, whereas the purple shaded low-bias regime reveals much richer conductance~modulations than a similar junction without the EuS~barrier~[recall~Fig.~\ref{FigCondThinBilayer}(a)]. }
                    \label{FigCondThinBilayerWithEuS}} 
         \end{figure}

            \hi{Furthermore, the conductance data of samples~B and C also reveal clearly visible zero-bias conductance~peaks, which we did not detect in~sample~A with the thicker $ \mathrm{Ni} $~($ 3 \, \mathrm{nm} $)--$ \mathrm{Ga} $~($ 60 \, \mathrm{nm} $)~bilayer. 
            Deducing the physical mechanism that causes such zero-bias~peaks is usually a highly nontrivial task, as they could stem from numerous distinct phenomena---like, for~example, zero-energy Andreev~reflections~\cite{Gueron1996,Pothier1997} or, as well \emph{in some cases}, superconducting triplet pairings~\cite{Kalcheim2015,Ouassou2017a}. 
            More specifically, a previous experimental study~\cite{Kalcheim2015} demonstrated that unexpectedly pronounced zero-bias conductance~peaks~(with amplitudes even exceeding those of their normal-state counterparts) arising in the tunneling~conductance of superconductor/half-metal~bilayers may be indicative of triplet superconductivity. 
            At the microscopic level, inhomogeneous magnetizations around the interface can flip some of the Cooper-pair electrons' spins, and thereby generate a ``mixture'' of spin-singlet and spin-triplet components in the superconducting order parameter~\cite{Bergeret2001b,Bergeret2001,Bergeret2001a,Eschrig2003,Houzet2007,Eschrig2008,Grein2009,Khaire2010,Robinson2010a,Robinson2010,Banerjee2014,Kalcheim2015,Ouassou2017a,Alidoust2018,Halterman2018,Diesch2018,Alidoust2020}. 
            }

            \hi{In bilayers that contain not fully spin-polarized ferromagnets~(i.e., no half~metals as the second electrode), the situation might be much more intriguing. 
            Thoroughly investigating electrical transport through Al/EuS~bilayers, Diesch~\textit{et~al.}~\cite{Diesch2018} revealed that---instead of a single zero-bias conductance~peak---superconducting triplet pairings give then rather rise to a rich, and not necessarily symmetric with respect to zero bias, low-bias double-peak conductance~pattern, in which the gap between the two newly forming conductance~peaks could be connected to the strength of the induced triplet pairings. 
            Within the applied STM~techniques, it was furthermore possible to individually address different transverse channels of the Al/EuS-bilayer junctions. 
            As a result, Diesch and coworkers proposed that the magnetization around the Al/EuS~interface is indeed inhomogeneous~(on a length scale of a few nanometers, which could refer to the grain~size of thin EuS~films), which is again most likely the mechanism that is responsible for the aforementioned ``mixing'' of singlet and triplet order parameters. 
            }

            \hi{Coming back to samples~B and C of our study, we must therefore conclude that the appearance of zero-bias conductance~peaks alone is neither a unique nor a sufficient signature of superconducting triplet pairings, and might as well originate from different physical effects. 
            However, as the main focus of our work is to understand the peculiar large-bias conductance~shoulders, and interpret these as clear fingerprints of triplet pairings, we did not further analyze the zero-bias peaks. 
            }

            \hi{As another remarkable experimental feature, our conductance~measurements on samples~B and C suggest that replacing the $ \mathrm{Ni} $~($ 1.6 \, \mathrm{nm} $)--$ \mathrm{Ga} $~($ 30 \, \mathrm{nm} $) by a $ \mathrm{Ga} $~($ 30 \, \mathrm{nm} $)--$ \mathrm{Ni} $~($ 1.6 \, \mathrm{nm} $)~bilayer---i.e., inverting the order of the Ni and Ga~films inside the bilayers---has no substantial effect on the large-bias conductance~shoulders~[see Figs.~\ref{FigCondThinBilayer}(a) and \ref{FigCondThinBilayer}(b)]. 
            We take this finding as an experimental hint that the physics being responsible for the formation of large-bias conductance~shoulders occurs \emph{around the Ni--Ga~interface}, and is thus quite independent of the order of the Ni and Ga~films. 
            Together with our claim that the conductance shoulders signify superconducting triplet pairings, we could therefore argue that the physical mechanism inducing triplet correlations in Ni--Ga~bilayers could be similar to the aforementioned ones; i.e., the magnetization around the interface of $ \mathrm{Ni} $~($ 1.6 \, \mathrm{nm} $)--$ \mathrm{Ga} $~($ 30 \, \mathrm{nm} $)~bilayers is inhomogeneous~(on a length scale of a few nanometers), which partially converts spin-singlet into spin-triplet Cooper~pairs through flipping some of the electrons' spins. 
            Nonetheless, this interpretation certainly requires more pertinent experimental evidence, which we will provide within our polarized~neutron~reflectometry measurements that are directly able to probe the profile of the interface~magnetizations and thereby support our predictions. 
            }

        \emph{Sample~D.} %%
            \hi{As the thin Ni~films of our junctions are only weakly ferromagnetic~(recall that we deduced spin~polarizations of about~$ 1 \, \% $ from Meservey--Tedrow spectroscopy)---and the triplet-pairing effects are thus most likely also rather moderate---, a more promising perspective could be to focus on samples that contain a second strongly ferromagnetic component. 
            The fourth studied sample consisted therefore of the $ \mathrm{Al} $~($ 4 \, \mathrm{nm} $)/$ \mathrm{Al}_2 \mathrm{O}_3 $~($ 0.4 \, \mathrm{nm} $)/$ \mathrm{EuS} $~($ 1.2 \, \mathrm{nm} $)/$ \mathrm{Ni} $~($ 1.6 \, \mathrm{nm} $)--$ \mathrm{Ga} $~($ 30 \, \mathrm{nm}) $/$ \mathrm{Al}_2 \mathrm{O}_3 $~($ 12 \, \mathrm{nm} $)~junction, which basically corresponds to sample~B except for the additional $ 1.2 \, \mathrm{nm} $-thick barrier composed of the strong ferromagnetic~insulator EuS~\cite{Moodera1988,Strambini2017}. 
            Such barriers have attracted considerable attention after earlier works~\cite{DeSimoni2018,Rouco2019a} had demonstrated that their high spin-filtering efficiency indeed provides an experimentally well-controllable way to convert more singlet into triplet Cooper~pairs, and thereby generate (almost) completely spin-polarized triplet supercurrents. 
            }

            The tunneling~conductance of sample~D at a temperature of about~$ 1 \, \mathrm{K} $ is presented in~Fig.~\ref{FigCondThinBilayerWithEuS}. 
            
            \hi{As the most important feature, we assert that the large-bias conductance shoulders---which we claimed to signify superconducting triplet pairings at the Ni--Ga~interface---are robust and even slightly more pronounced than in samples~B and C~(i.e., the splittings between the shoulders' conductance maxima are slightly larger). 
            This observation could be carefully interpreted as a possible experimental hint that the additional ferromagnetic EuS~spacer may indeed amplify the triplet pairings. 
            Moreover, adding EuS gives rise to extremely rich low-bias conductance~modulations that we could not detect in samples~A through C in which EuS was absent. 
            As we pointed out previously, STM~studies of Al/EuS~bilayers performed by Diesch~\textit{et~al.}~\cite{Diesch2018} indicated that low-bias conductance double peaks~(rather than just a single zero-bias conductance peak) could provide another signature of interfacial triplet pairings. 
            Our results obtained from sample~D look physically similar, apart from detecting four instead of two low-bias peaks~(two at negative and two at positive voltages, respectively). 
            Analogously to the zero-bias peaks of samples~B and C, unraveling the physical origin of these low-bias features has not yet been possible with the available data, and goes also beyond the scope of this manuscript. 
            One possible explanation for the doubling of low-bias peaks~(when compared to Diesch's work) might be that we are dealing with two different ferromagnetic films---EuS \emph{and} Ni---instead of just one~(EuS) as Diesch and coworkers, and thus need to consider two distinct interfaces at which the magnetization may be inhomogeneous. 
            However, this is a premature statement that we cannot uniquely confirm from our measurements. 
            }

    \subsection{Polarized~neutron~reflectometry}

        %% Fig. POLARIZED NEUTRON SPIN-ASYMMETRY RATIO %%
        \begin{figure}[tb]
            \centering
            \includegraphics[width=0.50\textwidth]{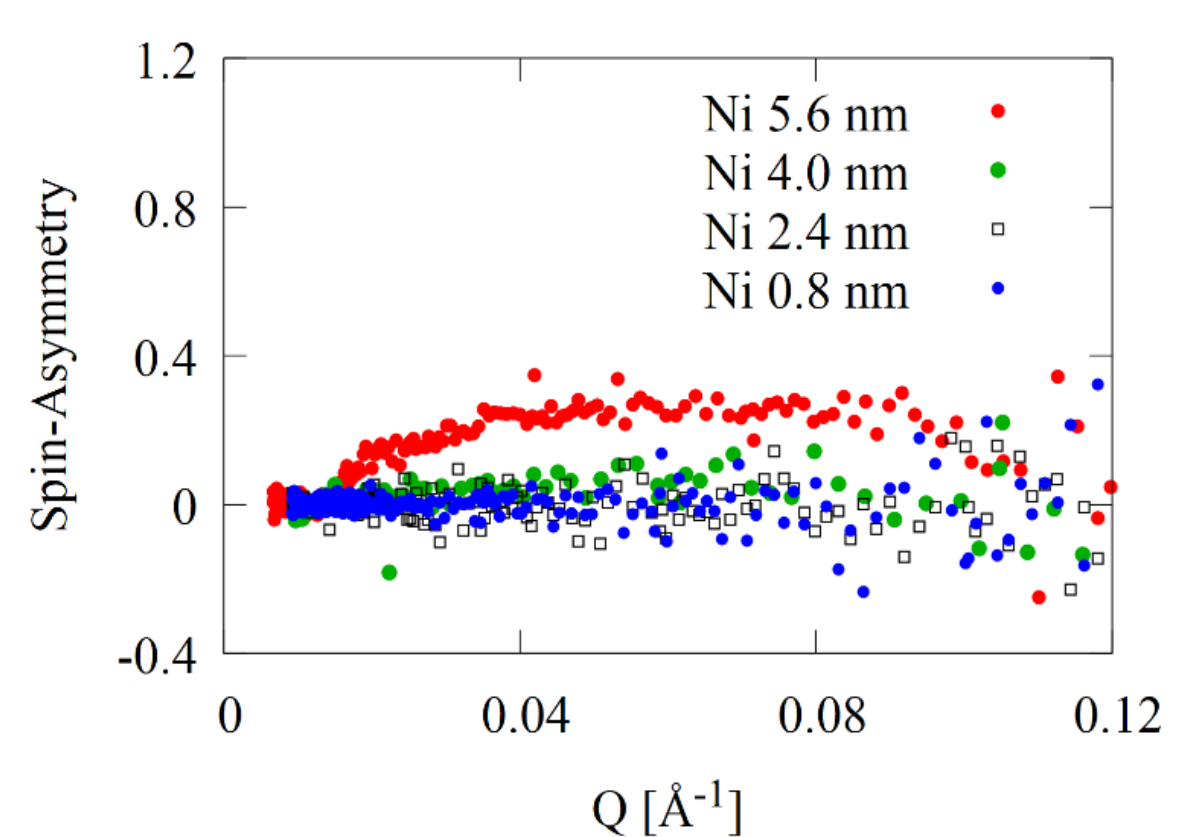}
            \caption{Polarized~neutron~reflectometry~(PNR) results for Ni--Ga~($ 25 \, \mathrm{nm} $)~bilayers with various Ni~thicknesses. 
                Spin-asymmetry~ratio~$ \mathrm{SA} = ({R^+} - {R^-}) / ({R^+} + {R^-}) $ obtained from the experimental reflectivity for spin-up~($ R^+ $) and spin-down~($ R^- $) neutron spin~states shown as a function of wave-vector~transfer~$ Q = 4\pi \sin \theta / \lambda $, where $ \theta $ indicates the incident angle and $ \lambda $ is the neutron wavelength. 
                \label{FigPolarizedNeutronSpinAsymmetry}}
        \end{figure}

        \hi{While discussing the tunneling-conductance~data obtained from samples~B and C in the preceding section, we argued that the observed large-bias conductance shoulders might serve as an experimentally accessible transport signature of superconducting triplet~pairings getting induced by inhomogeneous magnetizations around the interface of thinner $ \mathrm{Ni} $~($ 1.6 \, \mathrm{nm} $)--$ \mathrm{Ga} $~($ 30 \, \mathrm{nm}) $~bilayers. 
        In the following, we wish to provide \emph{independent} experimental evidence of this claim through a deeper characterization of the physical properties of this peculiar interface. }

        %% Fig. PNR and XRR %%
        \begin{figure}[tb]
            \centering
            \includegraphics[width=0.975\textwidth]{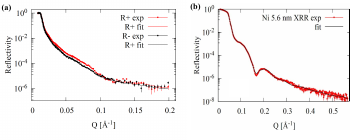}
            \caption{Polarized~neutron~reflectometry~(PNR) and x-ray~reflectometry~(XRR) results for the Ni~($ 5.6 \, \mathrm{nm} $)--Ga~($ 25 \, \mathrm{nm} $)~bilayer. 
            (a)~Experimental reflectivity as a function of the wave-vector~transfer~$ Q $ for spin-up~($ {R^+} $) and spin-down~($ {R^-} $) neutron spin states measured in~$ 0.1 \, \mathrm{T} $ after a zero-field~cooling to~$ 5 \, \mathrm{K} $. 
            (b)~Complementary XRR~data has been used to verify the films' depth~morphology. 
                The chemical and magnetization depth~profiles obtained from the fit to the data are shown in~Fig.~\ref{FigPNR}. 
                From the fit to the PNR and XRR, we deduce that the Ni~($ 5.6 \, \mathrm{nm} $)--Ga~($ 25 \, \mathrm{nm} $)~interface is sharp with a roughness of~$ 0.5 \, \mathrm{nm} $. 
                The Ga~layer's density is not uniform and consists of two roughly $ 7 \, \mathrm{nm} $- and $ 18 \, \mathrm{nm} $-thick sublayers~(see also Fig.~\ref{FigPNR}). 
                \label{FigPNRAndXRR}}
        \end{figure}

        For a deeper investigation of the Ni--Ga~interface, and to directly explore its structure and magnetization depth profile, we combine depth-sensitive polarized~neutron~reflectometry~(PNR) with low-angle x-ray reflectometry~(XRR)~studies. 
        Being electrically neutral, spin-polarized neutrons penetrate the entire multilayer junctions, probing the magnetic and structural composition of their films through buried interfaces down to the substrate~\cite{Lauter-Pasyuk2007}. 
        PNR allows for a direct determination of both the absolute value and the direction of the magnetic field induced inside the superconductor and was previously successfully applied to observe the diamagnetic Meissner~effect, as well as vortex-line distributions, in niobium- and YBCO-bilayer~films~\cite{Lauter-Pasyuk1998,Lauter-Pasyuk1999}. 
        In this paper, we report on the detection of \hi{\emph{inhomogeneous interface magnetizations} and } the \emph{paramagnetic Meissner~effect} in superconducting Ga, which \hi{altogether confirm our earlier claims that the proximity~coupling in the Ni--Ga~bilayer indeed induces superconducting triplet~states~\cite{Machida1978,Yokoyama2011a,Alidoust2014,Asano2014,DiBernardo2015a,Devizorova2019} near the interface, appearing to be responsible for the experimentally observed large-bias conductance~shoulders. }
        The PNR~experiments were performed on the Magnetism~Reflectometer at the Spallation~Neutron~Source at Oak~Ridge~National~Laboratory~\cite{Lauter-Pasyuk1998,Lauter-Pasyuk2009}, using a neutron~beam with a wavelength~band~$ \Delta \lambda $ of~$ 2.6 \, \text{\AA} $--$ 8.6 \, \text{\AA} $ and high polarization of~$ 98.5 \, \% $ to $ 99 \, \% $. 
        After cooling in zero field~(ZFC), measurements were conducted at temperatures of $ 15 \, \mathrm{K} $ and $ 5 \, \mathrm{K} $, with an external magnetic field applied in the plane of the sample up to $ 0.1 \, \mathrm{T} $. 
        Using the time-of-flight~method, a collimated polychromatic beam of polarized neutrons with a wavelength~band~$ \Delta \lambda $ impinges on the film at a grazing incidence~angle~$ \theta $, where it interacts with atomic nuclei and the spins of unpaired electrons. 
        Then, the reflected intensity is measured as a function of the wave-vector~transfer~$ Q = 4\pi \sin \theta / \lambda $ for two neutron~polarizations~$ {R^+} $ and $ {R^-} $ with the neutron~spin parallel~($ + $) or antiparallel~($ - $) to the direction of the external field~$ H_\mathrm{ext.} $; $ \lambda $ denotes the neutron wavelength. 
        To separate nuclear from magnetic scattering, we present our data in terms of the spin-asymmetry~ratio~$ \mathrm{SA} = ({R^+} - {R^-}) / ({R^+} + {R^-}) $. 
        For~example, a value of~$ \mathrm{SA} = 0 $ means that there is no magnetic moment in the system. 
        The depth~profiles of the \emph{nuclear} and \emph{magnetic scattering length densities}~(\emph{NSLD} and \emph{MSLD}) correspond to the depth~profiles of the chemical and in-plane magnetization vector distributions, respectively. 
        The total magnetization~$ M $ can be calculated from the MSLD~data using the relation~$ M (\mathrm{emu/cm^3}) = \mathrm{MSLD} (\text{\AA}^{-2}) / (2.853 \times 10^{-9}) $.

        %% Fig. PNR %%
        \begin{figure}[tb]
            \centering
            \includegraphics[width=0.975\textwidth]{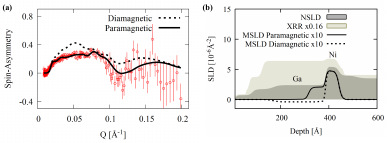}
            \caption{Polarized~neutron~reflectometry~(PNR) results for the Ni~($ 5.6 \, \mathrm{nm} $)--Ga~($ 25 \, \mathrm{nm} $)~bilayer. 
                (a)~PNR~spin-asymmetry~ratio~$ \mathrm{SA} = ({R^+} - {R^-}) / ({R^+} + {R^-}) $; the best fit to the data corresponds to the magnetization~profile shown as the solid black line in~(b)---i.e., to the \emph{paramagnetic Meissner~state}---, revealing $ 68 \, \mathrm{emu/cm^3} $ induced magnetization over about $ 7 \, \mathrm{nm} $ in Ga, while the Ni~film's magnetization is about $ 164 \, \mathrm{emu/cm^3} $ and uniform. 
                In contrast, the diamagnetic Meissner~state in the spin-asymmetry~(SA) plot and its magnetic scattering-length density~(MSLD) profile, referring to the dashed black lines, notably deviate from the experimental data. 
                    \label{FigPNR}}
        \end{figure}

        To verify the depth morphology of the films, we used complementary XRR~data.  
        These experiments were carried out on Ni--Ga~bilayers with the Ga~thickness fixed at about~$ 25 \, \mathrm{nm} $~(fitting our data, we obtain $ 21 \, \mathrm{nm} $~Ga~thickness), while the Ni~thickness was varied to cover $ 0.8 \, \mathrm{nm} $, $ 2.4 \, \mathrm{nm} $, $ 4.0 \, \mathrm{nm} $, and $ 5.6 \, \mathrm{nm} $; see~Fig.~\ref{FigPolarizedNeutronSpinAsymmetry}. 
        \hi{These bilayers are comparable~(in film thicknesses) to those in samples~B and C, which let the puzzling large-bias conductance shoulders occur. } 
        We explored the behavior of the magnetization of the bilayers above and below their critical temperature. 
        The samples were investigated under the same conditions as above, starting with ZFC down to~$ 15 \, \mathrm{K} $ and measuring at $ 0.1 \, \mathrm{T} $. 
        After that, the magnetic field was turned off, the sample was cooled to~$ 5 \, \mathrm{K} $, and the measurement was repeated at a magnetic field of~$ 0.1 \, \mathrm{T} $. 
        The sample with the $ 5.6 \, \mathrm{nm} $-thick Ni~film showed a clear magnetic signal~[i.e., clear SA~splitting between reflectivity for neutrons with spin~up~($ {R^+} $) and spin~down~($ {R^-} $)]. 
        For the sample containing the $ 4.0 \, \mathrm{nm} $-thick Ni~film, the SA magnetic~signal was reduced by a factor of~3, while no measurable magnetization could be detected~\hi{(within the accuracy of this method)} for the samples with the $ 2.4 \, \mathrm{nm} $- and $ 0.8 \, \mathrm{nm} $-thick Ni~films. 
        NSLD and MSLD depth profiles were obtained by simultaneous fitting to PNR and XRR~data~(shown in~Fig.~\ref{FigPNRAndXRR}), and finally plotted as a function of depth from the surface~(see Fig.~\ref{FigPNR}) for the sample with the $ 5.6 \, \mathrm{nm} $-thick Ni~film.

        %% Fig. PNR different Ni thickness %%
        \begin{figure}[tb]
            \centering
            \includegraphics[width=0.5\textwidth]{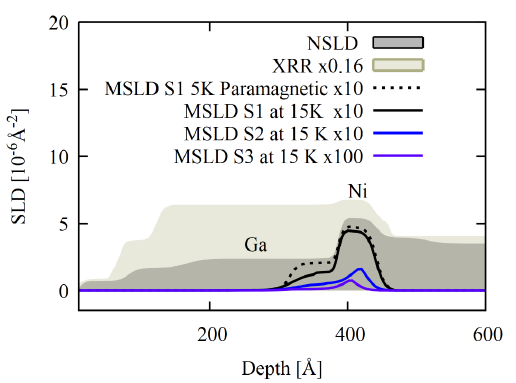}
            \caption{\hi{Polarized~neutron~reflectometry~(PNR) results~(see Fig.~\ref{FigPNR} for details) for Ni--Ga~($ 25 \, \mathrm{nm} $)~bilayers with $ 5.6 \, \mathrm{nm} $~(S1), $ 4 \, \mathrm{nm} $~(S2), and $ 2.4 \, \mathrm{nm} $~(S3) Ni~films. For $ 0.8 \, \mathrm{nm} $~Ni~(S4), we could not detect a net magnetization within the accuracy of our measurement. }
                    \label{FigPNRDifferentNiFilms}}
        \end{figure}

        To analyze the electromagnetic proximity~effect in Ni--Ga-bilayer structures from the PNR~data, we distinguish between two scenarios: (1)~the conventional \emph{diamagnetic Meissner~screening} and (2)~the \emph{paramagnetic Meissner~response} in Ga. 
        The results for both cases are shown in Fig.~\ref{FigPNR}. 
        Fitting the PNR~(obtained at~$ 5 \, \mathrm{K} $, which is below that bilayer's superconducting critical temperature) and XRR~data, we observe that the Ni~($ 5.6 \, \mathrm{nm} $)--Ga~($ 25 \, \mathrm{nm} $)~interface is sharp with a roughness of~$ 0.5 \, \mathrm{nm} $. 
        The thickness of the Ga~layer refined by the fit is~$ 21 \, \mathrm{nm} $. 
        The PNR spin-asymmetry~ratio~$ \mathrm{SA} = ({R^+} - {R^-}) / ({R^+} + {R^-}) $ reveals that the best fit to the data requires $ 68 \, \mathrm{emu/cm^3} $ induced magnetization over roughly $ 7 \, \mathrm{nm} $ in Ga in the vicinity of the Ni--Ga~interface, while the magnetization in the Ni~film is about $ 164 \, \mathrm{emu/cm^3} $ and uniform. 
        For comparison and confirmation of our findings, we additionally consider the model of \emph{diamagnetic Meissner~screening} with a penetrating flux. 
        In this case, the MSLD~profile will have a contribution from the magnetic field penetration~depth from both interfaces of the film. 
        Given that the Ga~layer is only $ 21 \, \mathrm{nm} $~thick, the magnetic field penetrates the entire film~\cite{Lauter-Pasyuk2000} so that the diamagnetic effect is significantly reduced~[see dashed line in~Fig.~\ref{FigPNR}(b)] and the corresponding SA in~Fig.~\ref{FigPNR}(a) shows considerable deviations from the experimental points. 
        We are thus able to directly see the induced ferromagnetic order's influence inside the Ga~layer \emph{right at the interface} with the Ni~film, which should be attributed to the \emph{paramagnetic Meissner~response}. 
        Therefore, PNR provides strong \hi{\emph{independent} experimental evidence~\cite{Machida1978,Yokoyama2011a,Alidoust2014,Asano2014,DiBernardo2015a} of our claims that inhomogeneous magnetizations induce superconducting triplet~pairings near the interface of the Ni--Ga~bilayer. }

        %% Fig. TB MODELING %%
        \begin{figure}[tb]
            \centering
            \includegraphics[width=0.975\textwidth]{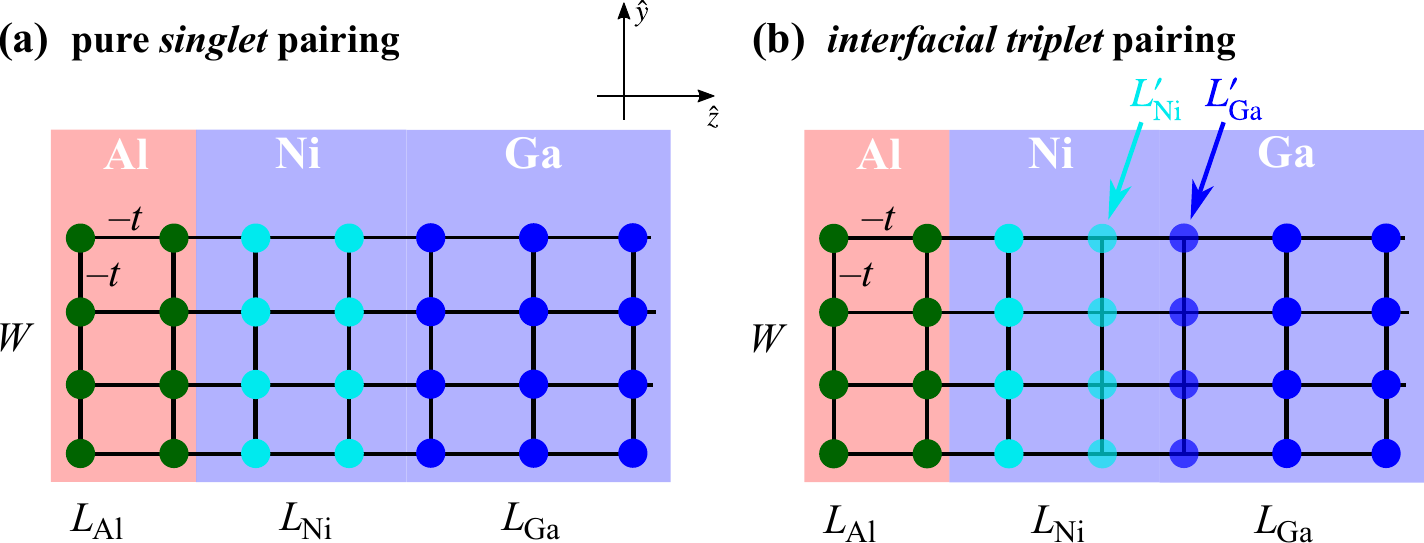}
            \caption{(a)~Schematic representation of the Al/Ni--Ga~junctions' tight-binding~modeling within the \textsc{Kwant}~Python transport~package, starting with a square~lattice with the constant~spacing~$ a = 1 \, [\text{arb. units}] $. 
                The numbers of horizontal lattice~sites inside the junctions' different regions are denoted by~$ L_\mathrm{Al} $, $ L_\mathrm{Ni} $, and~$ L_\mathrm{Ga} $, respectively, and that along the transverse direction by~$ W $. 
                The colored dots represent the \emph{on-site~energies}, determined by the Bogoliubov--de~Gennes Hamiltonian stated in~Eq.~(\ref{EqBdGKWANT}), while~$ t $ measures the strength of the \emph{nearest-neighbor hoppings}. 
                All superconducting gaps~(pairing~potentials) capture only pure singlet pairings. 
                (b)~Same as in~(a), but assuming that the superconducting gaps~(pairing~potentials) in Ni and Ga combine singlet with \emph{triplet} pairings within the $ L_\mathrm{Ni}' $ and $ L_\mathrm{Ga}' $ lattice~sites around their interface. 
                    \label{FigTBModeling}}
        \end{figure}

        \hi{To better illustrate the impact of the thickness of the Ni~film on the interface and magnetization~structure, we analyze the PNR~data obtained from Ni--Ga~bilayers with $ 4 \, \mathrm{nm} $, $ 2.4 \, \mathrm{nm} $, and $ 0.8 \, \mathrm{nm} $ Ni~films~(recall Fig.~\ref{FigPolarizedNeutronSpinAsymmetry}) in a similar manner; the thickness of Ga is still $ 25 \, \mathrm{nm} $. 
        From these data analyses, presented in~Fig.~\ref{FigPNRDifferentNiFilms}, we could indeed detect a significant difference between the structural~(nonmagnetic) and the magnetic Ni--Ga~interface roughness. 
        While the structural interfacial roughness is about~$ 0.5 \, \mathrm{nm} $ for $ 5.6 \, \mathrm{nm} $-thick Ni, we extract $ 1.5 \, \mathrm{nm} $ roughness for $ 4 \, \mathrm{nm} $-, $ 2.4 \, \mathrm{nm} $-, and $ 0.8 \, \mathrm{nm} $-thick Ni, respectively. 
        From the PNR~measurements performed at $ 0.1 \, \mathrm{mT} $ magnetic field and $ 15 \, \mathrm{K} $ temperature, we deduce that the interfacial magnetization is inhomogeneous and extending over several nanometers of $ 7 \, \mathrm{nm} $, $ 8 \, \mathrm{nm} $, and $ 9 \, \mathrm{nm} $~(for $ 5.6 \, \mathrm{nm} $-thick, $ 4 \, \mathrm{nm} $-thick, and $ 2.4 \, \mathrm{nm} $-thick Ni). 
        For the bilayer containing $ 0.8 \, \mathrm{nm} $ Ni, the accuracy of our approach does not suffice to resolve the magnetization profile. 
        As a more inhomogeneous interface magnetization is expected to convert more spin-singlet into spin-triplet Cooper~pairs to enhance the triplet-pairing mechanism, it could thus be promising to explore as a next step the tunneling~conductance of bilayers altering the thickness of the Ni~film~(and keeping Ga the same). 
        Increasing the Ni~thickness might then suppress and decreasing the Ni~thickness further amplify the triplet pairings, and therefore the conductance shoulders shall become either less or more pronounced. 
        }

        %%For completeness, we would like to mention that spin-polarized tunneling studies on Al/Al$ _2 $O$ _3 $/Ni--Ga tunnel~junctions, using Meservey--Tedrow~spectroscopy~\cite{Tedrow1971,Tedrow1973,Meservey1994} and a Zeeman-split superconducting Al spin~detector, revealed a very similar magnetic behavior as a function of the Ni--Ga~bilayers' Ni~thickness, i.e., for Ni~films thinner than $ 4 \, \mathrm{nm} $, it was hard to detect any spin~polarization~(could be as small as~$ 0.1 \, \% $), whereas a finite spin~polarization could be clearly measured for $ 4 \, \mathrm{nm} $ and larger Ni~thicknesses---being fully consistent with our PNR~observations. 
        %%These tunneling~measurements were performed at $ 0.48 \, \mathrm{K} $~temperature and with an applied in-plane magnetic~field of~$ 3.3 \, \mathrm{T} $. 

    \section{\hi{Theoretical toy model: Large-bias conductance~shoulders as signatures of interfacial triplet pairings} \label{Sec_Theory}}

        \hi{
        As we outlined in~Sec.~\ref{Sec_Exp} when discussing our experimental results, the appearance of large-bias shoulders in the tunneling~conductance of Ni--Ga~bilayers can serve as a signature of superconducting triplet~pairings induced at the Ni--Ga~interface---microscopically originating from inhomogeneous interface magnetizations, as our PNR~analyses clearly demonstrated. 
        In this final section of our paper, we formulate a purely phenomenological toy model to \emph{theoretically} convince \emph{at the qualitative level} that superconducting triplet pairings at the Ni--Ga~interface give indeed rise to the experimentally detected large-bias conductance shoulders. 
        }

        \subsection{\hi{Phenomenological toy model}}

        To describe quasiparticle excitations in \hi{Al/Ni--Ga~junctions~(for simplicity, our toy model neglects the Al$ _2 $O$ _3 $ tunneling~barriers)}, we formulate their  Bogoliubov--de~Gennes~Hamiltonian~\cite{DeGennes1989} on a lattice and compute the tunneling~DOS at zero temperature using the Python transport~package~\textsc{Kwant}~\cite{Groth2014}. 
        For simplicity, we consider a two-dimensional square~lattice with spacing~$ a = 1 \, [\text{arb. units}] $ between two adjacent lattice~sites; each site with the real-space~coordinates~$ (z,y) = (ai, aj) $ is uniquely identified by its integer lattice~indices~$ (i,j) $.  
        Figure~\ref{FigTBModeling}(a) shows a graphical representation of the chosen tight-binding~lattice. 
        We denote the numbers of lattice~sites along the longitudinal $ \hat{z} $-direction inside the \hi{Al, Ni, and Ga junction~regions by~$ L_\mathrm{Al} $, $ L_\mathrm{Ni} $, and~$ L_\mathrm{Ga} $,} respectively, whereas we assume in~total $ W $ lattice~sites along the transverse $ \hat{y} $-direction.

        The \emph{on-site~energies}~(with respect to the Fermi~level) at lattice~site~$ (i,j) $ are then given by the discretized Nambu-space Bogoliubov--de~Gennes Hamiltonian~[$ \Theta (\ldots) $ denotes the Heaviside step~function] 
		\hi{
		\allowdisplaybreaks	
		    \begin{align}
			    \hat{\mathcal{H}}_\mathrm{BdG} (i,j) &=
				    \Bigg[ 
				    4t \, \hat{\tau}_0 \nonumber \\[0.5 cm]
				    &\hspace{15 pt}+ \Delta_\mathrm{Al}^\mathrm{singlet} \, \hat{\tau}_2 \, \Theta (i) \, \Theta (L_\mathrm{Al} - i) \nonumber \\[0.5 cm]
				    &\hspace{15 pt}+ \Delta^\mathrm{XC}_\mathrm{Ni} \, \hat{\tau}_1 \, \Theta (i - L_\mathrm{Al} - 1) \, \Theta (L_\mathrm{Al} + L_\mathrm{Ni} - i) \nonumber \\[0.1 cm]
				    &\hspace{25 pt}+  \Delta_\text{Ni--Ga}^\mathrm{singlet} \, \hat{\tau}_2 \, \Theta (i - L_\mathrm{Al} - 1) \, \Theta (L_\mathrm{Al} + L_\mathrm{Ni} - i) \nonumber \\[0.1 cm]
				    &\hspace{25 pt}+  \Delta_\text{Ni--Ga}^\mathrm{triplet} \, \hat{\tau}_3 \, \Theta ( i - L_\mathrm{Al} - L_\mathrm{Ni} + L_\mathrm{Ni}' - 1 ) \, \Theta (L_\mathrm{Al} + L_\mathrm{Ni} - i) \nonumber \\[0.5 cm]
				    &\hspace{15 pt}+ \Delta_\text{Ni--Ga}^\mathrm{singlet} \, \hat{\tau}_2 \, \Theta (i - L_\mathrm{Al} - L_\mathrm{Ni} - 1) \, \Theta ( L_\mathrm{Al} + L_\mathrm{Ni} + L_\mathrm{Ga} - i ) \nonumber \\[0.1 cm]
				    &\hspace{25 pt}+  \Delta_\text{Ni--Ga}^\mathrm{triplet} \, \hat{\tau}_3 \, \Theta (i - L_\mathrm{Al} - L_\mathrm{Ni} - 1) \, \Theta \left( L_\mathrm{Al} + L_\mathrm{Ni} + L_\mathrm{Ga}' - i \right) \Bigg] \nonumber \\[0.5 cm]
				&\hspace{20 pt} \times \Theta(j) \, \Theta(W-1-j),
			\label{EqBdGKWANT}
		\end{align}
		\interdisplaylinepenalty=100000000
		}%%
		and the \emph{nearest-neighbor~hoppings}~($ \langle i, j \rangle $ indicates nearest-neighbor lattice~sites) by
		\begin{equation}
			\hat{\mathcal{H}}_\mathrm{hop} ( \langle i,j \rangle ) = -t \, \hat{\tau}_0 ,
			\label{EqHoppingKWANT}
		\end{equation}
		where
		\begin{align}
			\hat{\tau}_0 &= \left[ \begin{matrix} 1 & 0 & 0 & 0 \\ 0 & 1 & 0 & 0 \\ 0 & 0 & -1 & 0 \\ 0 & 0 & 0 & -1 \end{matrix} \right] ,\text{ } \hspace{10 pt} %%
			\hat{\tau}_1 = \left[ \begin{matrix} 1 & 0 & 0 & 0 \\ 0 & -1 & 0 & 0 \\ 0 & 0 & 1 & 0 \\ 0 & 0 & 0 & -1 \end{matrix} \right] ,\text{ } \hspace{10 pt} %%\\
			\hat{\tau}_2 = \left[ \begin{matrix} 0 & 0 & 1 & 0 \\ 0 & 0 & 0 & 1 \\ 1 & 0 & 0 & 0 \\ 0 & 1 & 0 & 0 \end{matrix} \right] ,  \nonumber \\[0.25 cm] %%
			\text{ and } \hspace{10 pt} 
			\hat{\tau}_3 &= \left[ \begin{matrix} 0 & 0 & 0 & 1 \\ 0 & 0 & 1 & 0 \\ 0 & 1 & 0 & 0 \\ 1 & 0 & 0 & 0 \end{matrix} \right] .
		\end{align}
		Thereby, the hopping~parameter represents~$ t = \hbar^2 / (2ma^2) $, where $ m $ refers to the effective quasiparticle~masses. 
		For our mostly to a qualitative level restricted modeling, it is most convenient to use such units that~$ t = 1 \, [\text{arb. units}] $.

		Apart from the discrete single-particle~energies~$ \varepsilon(i,j) = 4t \hat{\tau}_0 $ \hi{and the ferromagnetic exchange gap~$ \Delta^\mathrm{XC}_\mathrm{Ni} $ of Ni~(the magnetization~vector points along the $ \hat{z} $-direction),} we need to account for the films' distinct superconducting gaps. 
		The Bogoliubov--de~Gennes Hamiltonian involves now not only singlet superconducting gaps~(pairing~potentials), coupling spin-up and spin-down electrons to form \emph{spin-singlet Cooper~pairs}, but also triplet gaps~(pairing~potentials) that facilitate \emph{spin-triplet Cooper~pairs} consisting of two equal-spin electrons. 
		\hi{While the \emph{singlet} superconducting~gaps are abbreviated by~$ \Delta_\mathrm{Al}^\mathrm{singlet} $ and $ \Delta_\text{Ni--Ga}^\mathrm{singlet} $, the \emph{triplet} gap is denoted by~$ \Delta_\text{Ni--Ga}^\mathrm{triplet} $. }
		To ensure that triplet~correlations really only occur in the vicinity of the Ni--Ga~interface, their respective pairing-potential~terms are \emph{nonzero only} in $ L_\mathrm{Ni}' $ and~$ L_\mathrm{Ga}' $ of Ni's and Ga's lattice~sites around the Ni--Ga~interface, as we schematically illustrate in~Fig.~\ref{FigTBModeling}(b).

		Since the inhomogeneous magnetization at the Ni--Ga~interfaces, which ultimately induces the triplet pairings we are interested in, stems from a highly complex \hi{inhomogeneous} magnetic domain~structure that has not yet been fully understood at the microscopic level, we manually introduce the interfacial triplet~pairings into our toy model through including nonzero tunable equal-spin superconducting pairing~terms around the Ni--Ga~interface. 
		While this is sufficient to unravel the physical origin of the observed conductance~shoulders, 
		more comprehensive, and at the same time more realistic from a microscopic point of view, modeling is certainly desirable at a later stage, after gaining more experimental insight into the magnetic texture at the interface and possibly assisted by first-principles band-structure~calculations~\cite{Csire2018,Csire2018a, Gmitra2013}.

		To proceed, we implement the tight-binding Bogoliubov--de~Gennes Hamiltonian, given by~Eq.~\eqref{EqBdGKWANT}, in \textsc{Kwant}, and use \textsc{Kwant}'s internal Kernel~Polynomial~Method~(KPM) to extract the junctions' spatially integrated zero-temperature tunneling~DOS~(normalized to its normal-state counterpart) that we essentially probe through our conductance~measurements. 
		Along the transverse direction, we include $ W=500 $ lattice~sites. 
		Although changing~$ W $ does not \hi{qualitatively impact the tunneling~DOS}~(for that reason, we could also completely neglect the third spatial orientation, i.e., the $ \hat{x} $-direction, in our junctions), using rather large numbers is reasonable to minimize numerical~errors, which might cause unphysical numerical fluctuations in the DOS~data.

        \subsection{Tunneling-DOS simulations and large-bias conductance~shoulders}

\begin{figure}[tb]
            \centering
            \includegraphics[width=0.975\textwidth]{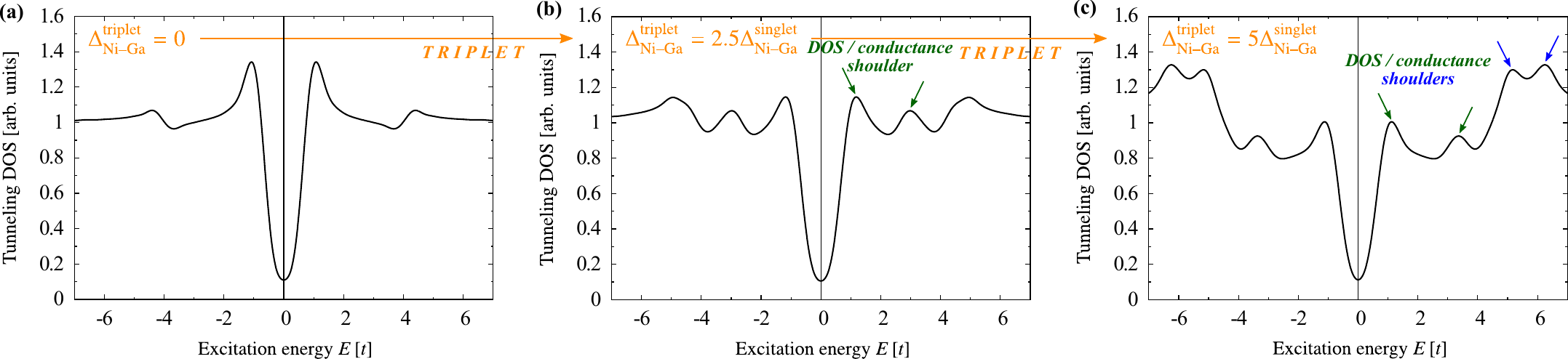}
            \caption{The \emph{calculated} zero-temperature tunneling~DOS of the Al~($ L_\mathrm{Al} = 400 $)/Ni~($ L_\mathrm{Ni} = 160 $)--Ga~($ L_\mathrm{Ga} = 3000 $)~junction (a)~\emph{in the absence of interfacial triplet~pairing}~($ \Delta_\text{Ni--Ga}^\mathrm{triplet} = 0 $), (b)~\emph{in the presence of moderate triplet~pairing}~($ \Delta_\text{Ni--Ga}^\mathrm{triplet} = 2.5 \Delta_\text{Ni--Ga}^\mathrm{singlet} $), and (c)~\emph{in the presence of strong triplet~pairing}~($ \Delta_\text{Ni--Ga}^\mathrm{triplet} = 5 \Delta_\text{Ni--Ga}^\mathrm{singlet} $) in $ L_\mathrm{Ni}' = L_\mathrm{Ni}/4 $ and $ L_\mathrm{Ga}' = L_\mathrm{Ga}/4 $ of the lattice~sites around the Ni--Ga~interface. 
            The particular choice of~$ L_\mathrm{Ni}' $ and $ L_\mathrm{Ga}' $ is not essential to reproduce the qualitative DOS~modulations, as long as $ L_\mathrm{Ni}' \ll L_\mathrm{Ni} $ and $ L_\mathrm{Ga}' \ll L_\mathrm{Ga} $.}
                \label{Fig_DOSDifferentTriplets}
        \end{figure}

        Figure~\ref{Fig_DOSDifferentTriplets} illustrates the computed tunneling~DOS of the Al/Ni--Ga~junction with the lattice-site~numbers~$ L_\mathrm{Al} = 400 $, $ L_\mathrm{Ni} = 160 $, and~$ L_\mathrm{Ga} = 3000 $; note that these were chosen such that their ratio~$ L_\mathrm{Al} : L_\mathrm{Ni} : L_\mathrm{Ga} = 400 : 160 : 3000 $ matches the film-thickness~ratio~$ d_\mathrm{Al} : d_\mathrm{Ni} : d_\mathrm{Ga} = 4 \, \mathrm{nm} : 1.6 \, \mathrm{nm} : 30 \, \mathrm{nm} $ of sample~B, although one cannot directly compare the theoretical and experimental dimensions as we chose $ a = 1 \, \text{[arb. units]} $ as the lattice~constant in our \textsc{Kwant}~simulations.  
        The specific values substituted for the singlet superconducting gaps of the Al~film and the Ni--Ga~bilayer are not particularly relevant to the results; we kept them at~$ \Delta_\mathrm{Al}^\mathrm{singlet} = 0.20t $ and~$ \Delta_\text{Ni--Ga}^\mathrm{singlet} = 0.80t $ in order to scale their ratio~$ \Delta_\text{Ni--Ga}^\mathrm{singlet} / \Delta_\mathrm{Al}^\mathrm{singlet} = 4 $ roughly according to the experimentally from sample~A extracted gaps. 
        The amplitude of the superconducting triplet gap, which is only present in the vicinity of the Ni--Ga~interface, was initially increased from~$ \Delta_\text{Ni--Ga}^\mathrm{triplet} = 0 $~(\emph{no triplet~pairing}) to~$ \Delta_\text{Ni--Ga}^\mathrm{triplet} = 2.5 \Delta_\text{Ni--Ga}^\mathrm{singlet} $~(\emph{moderate triplet~pairing}), and finally to~$ \Delta_\text{Ni--Ga}^\mathrm{triplet} = 5 \Delta_\text{Ni--Ga}^\mathrm{singlet} $~(\emph{strong triplet~pairing}). 
        Enhancing the triplet-pairing~strength in our model corresponds to produce a more inhomogeneously magnetized interface domain~texture in the experiment, which is expected to occur---coinciding with our experimental results---when decreasing the Ni(--Ga~bilayer)~thickness to get a rougher interface.

        Despite the particular values of triplet~gaps substituted into our calculations may seem exaggeratedly large when compared to their singlet counterparts, we can still use them in our numerical simulations to develop a \emph{qualitative} understanding of the underlying physics. 
        One of the reasons that such large triplet~gaps are required to reproduce the experimental findings could be the fact that we did not take care of formulating the full model using physically realistic units~(recall that we set, for~instance, $ a = 1 \, [\text{arb. units}]) $. 
        Furthermore, and as we mentioned earlier, the triplet-pairing~mechanisms in the real samples are a consequence of complex nonuniform interface magnetizations and are therefore at the microscopic level much more complicated to properly describe than in our strongly simplified phenomenological model. 
        The ``effectiveness'' of induced triplet~pairings depends additionally also on the strength of ferromagnetism inside the magnetic junction~films. 
        We assumed an extremely small exchange energy~gap in~Ni of~$ \Delta^\mathrm{XC}_\mathrm{Ni} = 0.5t $---when compared to the singlet superconducting gaps, which are typically orders of magnitude smaller than ferromagnetic exchange~couplings---in our simulations to demonstrate that even extremely weak ferromagnetism suffices to observe transport~ramifications of superconducting triplet~pairings. 
        Considering a typical metal Fermi~level~$ \mu \approx 10^3 \Delta_\text{Ni--Ga}^\mathrm{singlet} $, $ \Delta^\mathrm{XC}_\mathrm{Ni} = 0.5t $ corresponds to a Ni spin~polarization of just~$ \mathcal{P}_\mathrm{Ni} = (\Delta^\mathrm{XC}_\mathrm{Ni} / 2) / \mu \approx 0.03 \, \% $. 
        Although Meservey--Tedrow~spectroscopy indicated that the Ni~spin~polarization in our samples \hi{is only about~$ 1 \, \% $, this is still notably larger than the tiny value assumed for our simulations}, which could provide another reason for the large effective triplet~pairings required in our simulations to recover the experimentally evident features.

        In the absence of triplet~pairing at the Ni--Ga~interface, we recover two DOS~maxima~(becoming visible in our transport~measurements in terms of conductance~maxima) that we associated with the energies of the singlet superconducting~gaps of Al and the Ni--Ga~bilayer when analyzing the experimental~data obtained from sample~A~(recall the Rowell-Feldman~description explained in~Sec.~\ref{Sec_Exp}). 
        Already moderate triplet~pairing, however, turns the first DOS~maximum at the lower of the two bias~voltages into two neighboring maxima, and thereby forms the first shoulder that we probably witnessed in the conductance~spectrum of sample~B. 
        The second experimentally observed shoulder at slightly larger bias~voltage eventually appears when the triplet-pairing~strength is further enhanced. 
        The latter shoulder~(i.e., the one at larger bias~voltage) seems to be less sensitive to interfacial triplet pairings in~general since the splitting between its two neighboring conductance~maxima is substantially smaller than that within the first---in good agreement with our experimental results~(recall the conductance~data of samples~B and C presented in~Fig.~\ref{FigCondThinBilayer})---and requires thus a more sizable triplet-pairing~strength to become indeed evident. 
        \hi{\emph{Although our theoretical DOS~simulations are robust enough to qualitatively demonstrate that the experimentally detected large-bias conductance~shoulders indeed provide a transport~fingerprint of triplet~pairings at the Ni--Ga~bilayer interface, theory and experiment can, at least at this point, not be compared to each other at the quantitative level due to the strong simplifications made in our model.} }

    \section{Conclusions    \label{Sec_Conclusions}}

        In summary, we thoroughly discussed and analyzed our tunneling-conductance~measurements on superconducting magnetic Al/$ \mathrm{Al}_2 \mathrm{O}_3 $(/EuS)/Ni--Ga~junctions, focusing, in~particular, on thicker Ni~($ 3 \, \mathrm{nm} $)--Ga~($ 60 \, \mathrm{nm} $) and thinner $ \mathrm{Ni} $~($ 1.6 \, \mathrm{nm} $)--$ \mathrm{Ga} $~($ 30 \, \mathrm{nm} $)~bilayers, respectively. 
        While the conductance~spectrum in the first case could be explained based on the findings of earlier studies, the second scenario turned out to become much more puzzling, as it mainly led to the additional formation of unexpected large-bias conductance~shoulders that have not yet been understood. 
        Since the latter remained mostly unaffected when ``inverting'' the Ni and Ga~films, we concluded that all important physics should happen near the Ni--Ga~interface.

        Performing PNR~analyses to collect more information about the structure and magnetization of this interface, we detected the paramagnetic Meissner~response in~Ga \hi{to convince } that the proximity-coupled bilayer \hi{induces} superconducting triplet~pairings around the Ni--Ga~interface. 
        With this in mind, we elaborated on a strongly simplified theoretical \hi{toy model, which } allowed us to compute the junctions' tunneling~DOS that our conductance~measurements essentially probe. 
        Comparing our phenomenological DOS~simulations with experimental conductance~data \hi{substantiated} that the conductance~shoulders do indeed provide a well-accessible transport~fingerprint of newly induced superconducting triplet~correlations in the vicinity of the Ni--Ga~interface.

        To further characterize the novel triplet~pairings within the Ni--Ga~bilayer, we suggest to subsequently analyze the Ni--Ga-interface~profile through SQUID and Lorentz-microscopy measurements, which can directly probe inhomogeneous spin~textures around the interface. 
        \hi{Moreover, investigating our samples' transport~characteristics in the presence of an external magnetic field might give deeper insight into the triplet-pairing~mechanism, as this manipulates the inhomogeneity of the interface magnetizations and shall therefore give a unique magnetization dependence to the conductance~shoulders. }

    %% Acknowledgments %%
        \ack
        The experimental work performed in the U.S. was supported by the NSF C-Accel. Track~C under Grant No.~2040620, NSF Grant DMR~1700137, ONR Grants N00014-16-1-2657 and~N00014-20-1-2306, and John~Templeton~Foundation Grants~39944 and~60148. 
        The theoretical work at the University of Regensburg~(A.C. and J.F.) received funding from the Elite~Network of Bavaria through the International~Doctorate~Program Topological~Insulators, and Deutsche Forschungsgemeinschaft~(DFG, German Research Foundation) through Subproject~B07 within the Collaborative~Research~Center SFB~1277~(Project-ID~314695032) and the Research~Grant ``Spin and magnetic properties of superconducting tunnel~junctions''~(Project-ID~454646522). 
        The undergraduate M.S. was supported by the UROP~program~funds at Massachusetts Institute of Technology. 
        This research used resources at the Spallation~Neutron~Source, a DOE~Office of Science~User~Facility operated by the Oak~Ridge~National~Laboratory.

    %% References %%
        \section*{References}
    
        \bibliographystyle{iopart_num}
        \bibliography{paper}

\providecommand{\newblock}{}
\begin{thebibliography}{100}
\expandafter\ifx\csname url\endcsname\relax
  \def\url#1{{\tt #1}}\fi
\expandafter\ifx\csname urlprefix\endcsname\relax\def\urlprefix{URL }\fi
\providecommand{\eprint}[2][]{\url{#2}}

\bibitem{Fabian2004}
{\v{Z}}uti{\'{c}} I, Fabian J and {Das Sarma} S 2004 {\em Rev. Mod. Phys.\/}
  {\bf 76} 323--410
  \urlprefix\url{http://link.aps.org/doi/10.1103/RevModPhys.76.323}

\bibitem{Fabian2007}
Fabian J, Matos-Abiague A, Ertler C, Stano P and {\v{Z}}uti{\'{c}} I 2007 {\em
  Acta Phys. Slovaca\/} {\bf 57} 565--907
  \urlprefix\url{http://www.physics.sk/aps/pub.php?y=2007&pub=aps-07-04}

\bibitem{Eschrig2011}
Eschrig M 2011 {\em Phys. Today\/} {\bf 64} 43--49
  \urlprefix\url{http://scitation.aip.org/content/aip/magazine/physicstoday/article/64/1/10.1063/1.3541944}

\bibitem{Linder2015}
Linder J and Robinson J~W~A 2015 {\em Sci. Rep.\/} {\bf 5} 15483 ISSN 2045-2322
  \urlprefix\url{http://dx.doi.org/10.1038/srep15483
  http://www.nature.com/articles/srep15483}

\bibitem{Ohnishi2020}
Ohnishi K, Komori S, Yang G, Jeon K~R, {Olde Olthof} L~A~B, Montiel X, Blamire
  M~G and Robinson J~W~A 2020 {\em Appl. Phys. Lett.\/} {\bf 116} 130501 ISSN
  0003-6951 \urlprefix\url{http://aip.scitation.org/doi/10.1063/1.5138905}

\bibitem{Ioffe1999}
Ioffe L~B, Geshkenbein V~B, Feigel'man M~V, Fauch{\`e}re A~L and Blatter G 1999
  {\em Nature\/} {\bf 398} 679--681
  \urlprefix\url{http://www.nature.com/nature/journal/v398/n6729/abs/398679a0.html}

\bibitem{Mooij1999}
Mooij J~E, Orlando T~P, Levitov L, Tian L, van~der Wal C~H and Lloyd S 1999
  {\em Science\/} {\bf 285} 1036--1039
  \urlprefix\url{http://science.sciencemag.org/content/285/5430/1036}

\bibitem{Blatter2001}
Blatter G, Geshkenbein V~B and Ioffe L~B 2001 {\em Phys. Rev. B\/} {\bf 63}
  174511 ISSN 0163-1829
  \urlprefix\url{https://link.aps.org/doi/10.1103/PhysRevB.63.174511}

\bibitem{Ustinov2003}
Ustinov A~V and Kaplunenko V~K 2003 {\em J. Appl. Phys.\/} {\bf 94} 5405 ISSN
  00218979
  \urlprefix\url{http://scitation.aip.org/content/aip/journal/jap/94/8/10.1063/1.1604964}

\bibitem{Yamashita2005}
Yamashita T, Tanikawa K, Takahashi S and Maekawa S 2005 {\em Phys. Rev.
  Lett.\/} {\bf 95} 097001
  \urlprefix\url{http://link.aps.org/doi/10.1103/PhysRevLett.95.097001}

\bibitem{Feofanov2010}
Feofanov A~K, Oboznov V~A, Bol'ginov V~V, Lisenfeld J, Poletto S, Ryazanov V~V,
  Rossolenko A~N, Khabipov M, Balashov D, Zorin A~B, Dmitriev P~N, Koshelets
  V~P and Ustinov A~V 2010 {\em Nat. Phys.\/} {\bf 6} 593--597
  \urlprefix\url{http://www.nature.com/doifinder/10.1038/nphys1700}

\bibitem{Khabipov2010}
Khabipov M~I, Balashov D~V, Maibaum F, Zorin A~B, Oboznov V~A, Bolginov V~V,
  Rossolenko A~N and Ryazanov V~V 2010 {\em Supercond. Sci. Technol.\/} {\bf
  23} 045032 ISSN 0953-2048
  \urlprefix\url{https://iopscience.iop.org/article/10.1088/0953-2048/23/4/045032}

\bibitem{Devoret2013}
Devoret M~H and Schoelkopf R~J 2013 {\em Science\/} {\bf 339} 1169--1174
  \urlprefix\url{http://science.sciencemag.org/content/339/6124/1169}

\bibitem{Soulen1998}
{Soulen Jr} R~J 1998 {\em Science\/} {\bf 282} 85--88 ISSN 00368075
  \urlprefix\url{https://www.sciencemag.org/lookup/doi/10.1126/science.282.5386.85}

\bibitem{Soulen1999}
{Soulen Jr} R~J, Osofsky M~S, Nadgorny B, Ambrose T, Broussard P, Cheng S~F,
  Byers J, Tanaka C~T, Nowack J, Moodera J~S, Laprade G, Barry A and Coey M~D
  1999 {\em J. Appl. Phys.\/} {\bf 85} 4589--4591 ISSN 00218979
  \urlprefix\url{http://scitation.aip.org/content/aip/journal/jap/85/8/10.1063/1.370417}

\bibitem{DeJong1995}
{de Jong} M~J~M and Beenakker C~W~J 1995 {\em Phys. Rev. Lett.\/} {\bf 74}
  1657--1660
  \urlprefix\url{http://link.aps.org/doi/10.1103/PhysRevLett.74.1657}

\bibitem{Golubov2004}
Golubov A~A, Kupriyanov M~Y and Il'ichev E 2004 {\em Rev. Mod. Phys.\/} {\bf
  76} 411--469
  \urlprefix\url{http://link.aps.org/doi/10.1103/RevModPhys.76.411}

\bibitem{Yu1965}
Yu L 1965 {\em Acta Phys. Sin.\/} {\bf 21} 75--91
  \urlprefix\url{http://wulixb.iphy.ac.cn/CN/Y1965/V21/I1/75}

\bibitem{Shiba1968}
Shiba H 1968 {\em Prog. Theor. Phys.\/} {\bf 40} 435--451
  \urlprefix\url{https://academic.oup.com/ptp/article/40/3/435/1831894/Classical-Spins-in-Superconductors}

\bibitem{Rusinov1968}
Rusinov A~I 1968 {\em Zh. Eksp. Teor. Fiz. Pisma Red.\/} {\bf 9} 146

\bibitem{Rusinov1968alt}
Rusinov A~I 1969 {\em JETP Lett.\/} {\bf 9} 85
  \urlprefix\url{http://www.jetpletters.ac.ru/ps/1658/article_25295.shtml}

\bibitem{Costa2018}
Costa A, Fabian J and Kochan D 2018 {\em Phys. Rev. B\/} {\bf 98} 134511 ISSN
  2469-9950 \urlprefix\url{http://arxiv.org/abs/1803.02063
  https://link.aps.org/doi/10.1103/PhysRevB.98.134511}

\bibitem{Kochan2020}
Kochan D, Barth M, Costa A, Richter K and Fabian J 2020 {\em Phys. Rev.
  Lett.\/} {\bf 125} 087001 ISSN 1079-7114
  \urlprefix\url{https://doi.org/10.1103/PhysRevLett.125.087001}

\bibitem{Bulaevskii1977a}
Bulaevskii L~N, Kuzii V~V and Sobyanin A~A 1977 {\em Pis'ma Zh. Eksp. Teor.
  Fiz.\/} {\bf 25} 314--318

\bibitem{Bulaevskii1977b}
Bulaevskii L~N, Kuzii V~V and Sobyanin A~A 1977 {\em JETP Lett.\/} {\bf 25}
  290--294
  \urlprefix\url{http://www.jetpletters.ac.ru/ps/1410/article_21163.shtml}

\bibitem{Ryazanov2001}
Ryazanov V~V, Oboznov V~A, Rusanov A~Y, Veretennikov A~V, Golubov A~A and Aarts
  J 2001 {\em Phys. Rev. Lett.\/} {\bf 86} 2427--2430
  \urlprefix\url{http://link.aps.org/doi/10.1103/PhysRevLett.86.2427}

\bibitem{Nilsson2008}
Nilsson J, Akhmerov A~R and Beenakker C~W~J 2008 {\em Phys. Rev. Lett.\/} {\bf
  101} 120403 ISSN 0031-9007
  \urlprefix\url{https://link.aps.org/doi/10.1103/PhysRevLett.101.120403}

\bibitem{Duckheim2011}
Duckheim M and Brouwer P~W 2011 {\em Phys. Rev. B\/} {\bf 83} 054513
  \urlprefix\url{http://link.aps.org/doi/10.1103/PhysRevB.83.054513}

\bibitem{Lee2012a}
Lee S~P, Alicea J and Refael G 2012 {\em Phys. Rev. Lett.\/} {\bf 109} 126403
  ISSN 0031-9007
  \urlprefix\url{https://link.aps.org/doi/10.1103/PhysRevLett.109.126403}

\bibitem{Nadj-Perge2014}
Nadj-Perge S, Drozdov I~K, Li J, Chen H, Jeon S, Seo J, MacDonald A~H, Bernevig
  B~A and Yazdani A 2014 {\em Science\/} {\bf 346} 602--607
  \urlprefix\url{http://www.sciencemag.org/content/346/6209/602.abstract}

\bibitem{Dumitrescu2015}
Dumitrescu E, Roberts B, Tewari S, Sau J~D and {Das Sarma} S 2015 {\em Phys.
  Rev. B\/} {\bf 91} 094505 ISSN 1098-0121
  \urlprefix\url{https://link.aps.org/doi/10.1103/PhysRevB.91.094505}

\bibitem{Pawlak2016}
Pawlak R, Kisiel M, Klinovaja J, Meier T, Kawai S, Glatzel T, Loss D and Meyer
  E 2016 {\em npj Quantum Inf.\/} {\bf 2} 16035 ISSN 2056-6387
  \urlprefix\url{http://www.nature.com/articles/npjqi201635}

\bibitem{Ruby2017}
Ruby M, Heinrich B~W, Peng Y, von Oppen F and Franke K~J 2017 {\em Nano
  Lett.\/} {\bf 17} 4473--4477 ISSN 1530-6984
  \urlprefix\url{http://pubs.acs.org/doi/10.1021/acs.nanolett.7b01728}

\bibitem{Livanas2019}
Livanas G, Sigrist M and Varelogiannis G 2019 {\em Sci. Rep.\/} {\bf 9} 6259
  ISSN 2045-2322
  \urlprefix\url{http://www.nature.com/articles/s41598-019-42558-3}

\bibitem{Manna2020}
Manna S, Wei P, Xie Y, Law K~T, Lee P~A and Moodera J~S 2020 {\em Proc. Natl.
  Acad. Sci.\/} {\bf 117} 8775--8782 ISSN 0027-8424
  \urlprefix\url{http://www.pnas.org/lookup/doi/10.1073/pnas.1919753117}

\bibitem{Hoegl2015}
H{\"{o}}gl P, Matos-Abiague A, {\v{Z}}uti{\'{c}} I and Fabian J 2015 {\em Phys.
  Rev. Lett.\/} {\bf 115} 116601
  \urlprefix\url{http://journals.aps.org/prl/abstract/10.1103/PhysRevLett.115.116601}

\bibitem{Hoegl2015a}
H{\"{o}}gl P, Matos-Abiague A, {\v{Z}}uti{\'{c}} I and Fabian J 2015 {\em Phys.
  Rev. Lett.\/} {\bf 115} 159902(E)
  \urlprefix\url{https://link.aps.org/doi/10.1103/PhysRevLett.115.159902}

\bibitem{Jacobsen2016}
Jacobsen S~H, Kulagina I and Linder J 2016 {\em Sci. Rep.\/} {\bf 6} 23926 ISSN
  2045-2322 \urlprefix\url{http://www.nature.com/articles/srep23926}

\bibitem{Costa2017}
Costa A, H\"ogl P and Fabian J 2017 {\em Phys. Rev. B\/} {\bf 95}(2) 024514
  \urlprefix\url{https://link.aps.org/doi/10.1103/PhysRevB.95.024514}

\bibitem{Martinez2020}
Mart{\'{i}}nez I, H{\"{o}}gl P, Gonz{\'{a}}lez-Ruano C, Cascales J~P, Tiusan C,
  Lu Y, Hehn M, Matos-Abiague A, Fabian J, {\v{Z}}uti{\'{c}} I and Aliev F~G
  2020 {\em Phys. Rev. Appl.\/} {\bf 13} 014030 ISSN 2331-7019
  \urlprefix\url{https://link.aps.org/doi/10.1103/PhysRevApplied.13.014030}

\bibitem{Keizer2006}
Keizer R~S, Goennenwein S~T~B, Klapwijk T~M, Miao G, Xiao G and Gupta A 2006
  {\em Nature\/} {\bf 439} 825--827
  \urlprefix\url{http://www.nature.com/doifinder/10.1038/nature04499}

\bibitem{Moodera1990}
Moodera J~S and Meservey R 1990 {\em Phys. Rev. B\/} {\bf 42} 179--183 ISSN
  0163-1829 \urlprefix\url{https://link.aps.org/doi/10.1103/PhysRevB.42.179}

\bibitem{LeClair2005}
LeClair P, Moodera J~S, Philip J and Heiman D 2005 {\em Phys. Rev. Lett.\/}
  {\bf 94} 037006 ISSN 0031-9007
  \urlprefix\url{http://link.aps.org/doi/10.1103/PhysRevLett.94.037006
  https://link.aps.org/doi/10.1103/PhysRevLett.94.037006}

\bibitem{Belzig1996}
Belzig W, Bruder C and Sch{\"{o}}n G 1996 {\em Phys. Rev. B\/} {\bf 53}
  5727--5733 ISSN 0163-1829
  \urlprefix\url{https://link.aps.org/doi/10.1103/PhysRevB.53.5727}

\bibitem{Kadigrobov2001}
Kadigrobov A, Shekhter R~I and Jonson M 2001 {\em Europhys. Lett.\/} {\bf 54}
  394--400 ISSN 0295-5075
  \urlprefix\url{https://iopscience.iop.org/article/10.1209/epl/i2001-00107-2}

\bibitem{Bergeret2001}
Bergeret F~S, Volkov A~F and Efetov K~B 2001 {\em Phys. Rev. Lett.\/} {\bf 86}
  4096--4099
  \urlprefix\url{http://link.aps.org/doi/10.1103/PhysRevLett.86.4096}

\bibitem{Bergeret2005}
Bergeret F~S, Volkov A~F and Efetov K~B 2005 {\em Rev. Mod. Phys.\/} {\bf 77}
  1321--1373 \urlprefix\url{http://link.aps.org/doi/10.1103/RevModPhys.77.1321}

\bibitem{Yokoyama2007a}
Yokoyama T, Tanaka Y and Golubov A~A 2007 {\em Phys. Rev. B\/} {\bf 75} 134510
  ISSN 1098-0121
  \urlprefix\url{https://link.aps.org/doi/10.1103/PhysRevB.75.134510}

\bibitem{Linder2009}
Linder J, Yokoyama T, Sudb{\o} A and Eschrig M 2009 {\em Phys. Rev. Lett.\/}
  {\bf 102} 107008 ISSN 0031-9007
  \urlprefix\url{https://link.aps.org/doi/10.1103/PhysRevLett.102.107008}

\bibitem{Khaire2010}
Khaire T~S, Khasawneh M~A, {Pratt Jr} W~P and Birge N~O 2010 {\em Phys. Rev.
  Lett.\/} {\bf 104} 137002
  \urlprefix\url{https://link.aps.org/doi/10.1103/PhysRevLett.104.137002}

\bibitem{Robinson2010}
Robinson J~W~A, Witt J~D~S and Blamire M~G 2010 {\em Science\/} {\bf 329}
  59--61 ISSN 0036-8075
  \urlprefix\url{http://science.sciencemag.org/content/329/5987/59}

\bibitem{Anwar2010}
Anwar M~S, Czeschka F, Hesselberth M, Porcu M and Aarts J 2010 {\em Phys. Rev.
  B\/} {\bf 82} 100501(R) ISSN 1098-0121
  \urlprefix\url{https://link.aps.org/doi/10.1103/PhysRevB.82.100501}

\bibitem{Yokoyama2011a}
Yokoyama T, Tanaka Y and Nagaosa N 2011 {\em Phys. Rev. Lett.\/} {\bf 106}
  246601 ISSN 0031-9007
  \urlprefix\url{https://link.aps.org/doi/10.1103/PhysRevLett.106.246601}

\bibitem{Bergeret2012}
Bergeret F~S, Verso A and Volkov A~F 2012 {\em Phys. Rev. B\/} {\bf 86}
  060506(R) ISSN 1098-0121
  \urlprefix\url{https://link.aps.org/doi/10.1103/PhysRevB.86.060506}

\bibitem{Bergeret2013}
Bergeret F~S and Tokatly I~V 2013 {\em Phys. Rev. Lett.\/} {\bf 110} 117003
  \urlprefix\url{http://link.aps.org/doi/10.1103/PhysRevLett.110.117003}

\bibitem{Bergeret2014}
Bergeret F~S and Tokatly I~V 2014 {\em Phys. Rev. B\/} {\bf 89} 134517
  \urlprefix\url{http://link.aps.org/doi/10.1103/PhysRevB.89.134517}

\bibitem{Alidoust2015a}
Alidoust M, Halterman K and Valls O~T 2015 {\em Phys. Rev. B\/} {\bf 92} 014508
  ISSN 1098-0121
  \urlprefix\url{https://link.aps.org/doi/10.1103/PhysRevB.92.014508}

\bibitem{DiBernardo2015}
{Di Bernardo} A, Diesch S, Gu Y, Linder J, Divitini G, Ducati C, Scheer E,
  Blamire M~G and Robinson J~W~A 2015 {\em Nat. Commun.\/} {\bf 6} 8053 ISSN
  2041-1723 \urlprefix\url{http://www.nature.com/articles/ncomms9053}

\bibitem{Arjoranta2016}
Arjoranta J and Heikkil{\"{a}} T~T 2016 {\em Phys. Rev. B\/} {\bf 93} 024522
  ISSN 2469-9950
  \urlprefix\url{https://link.aps.org/doi/10.1103/PhysRevB.93.024522}

\bibitem{Espedal2016}
Espedal C, Yokoyama T and Linder J 2016 {\em Phys. Rev. Lett.\/} {\bf 116}
  127002 ISSN 0031-9007
  \urlprefix\url{https://link.aps.org/doi/10.1103/PhysRevLett.116.127002}

\bibitem{Pal2017}
Pal A, Ouassou J~A, Eschrig M, Linder J and Blamire M~G 2017 {\em Sci. Rep.\/}
  {\bf 7} 3--7 ISSN 20452322
  \urlprefix\url{http://dx.doi.org/10.1038/srep40604}

\bibitem{Bergeret2020}
Bergeret F~S and Tokatly I~V 2020 {\em Phys. Rev. B\/} {\bf 102} 060506(R) ISSN
  2469-9950
  \urlprefix\url{https://link.aps.org/doi/10.1103/PhysRevB.102.060506}

\bibitem{Bergeret2001b}
Bergeret F~S, Volkov A~F and Efetov K~B 2001 {\em Phys. Rev. Lett.\/} {\bf 86}
  3140--3143 ISSN 0031-9007
  \urlprefix\url{https://link.aps.org/doi/10.1103/PhysRevLett.86.3140}

\bibitem{Bergeret2001a}
Bergeret F~S, Volkov A~F and Efetov K~B 2001 {\em Phys. Rev. B\/} {\bf 64}
  134506 \urlprefix\url{http://link.aps.org/doi/10.1103/PhysRevB.64.134506}

\bibitem{Eschrig2003}
Eschrig M, Kopu J, Cuevas J~C and Sch{\"{o}}n G 2003 {\em Phys. Rev. Lett.\/}
  {\bf 90} 137003 ISSN 0031-9007
  \urlprefix\url{https://link.aps.org/doi/10.1103/PhysRevLett.90.137003}

\bibitem{Houzet2007}
Houzet M and Buzdin A~I 2007 {\em Phys. Rev. B\/} {\bf 76} 060504(R) ISSN
  1098-0121 \urlprefix\url{https://link.aps.org/doi/10.1103/PhysRevB.76.060504}

\bibitem{Eschrig2008}
Eschrig M and L{\"{o}}fwander T 2008 {\em Nat. Phys.\/} {\bf 4} 138--143
  \urlprefix\url{http://dx.doi.org/10.1038/nphys831}

\bibitem{Grein2009}
Grein R, Eschrig M, Metalidis G and Sch{\"{o}}n G 2009 {\em Phys. Rev. Lett.\/}
  {\bf 102} 227005 ISSN 0031-9007
  \urlprefix\url{https://link.aps.org/doi/10.1103/PhysRevLett.102.227005}

\bibitem{Robinson2010a}
Robinson J~W~A, Hal\'asz G~B, Buzdin A~I and Blamire M~G 2010 {\em Phys. Rev.
  Lett.\/} {\bf 104} 207001
  \urlprefix\url{https://link.aps.org/doi/10.1103/PhysRevLett.104.207001}

\bibitem{Banerjee2014}
Banerjee N, Robinson J~W~A and Blamire M~G 2014 {\em Nat. Commun.\/} {\bf 5}
  1--6 ISSN 20411723

\bibitem{Diesch2018}
Diesch S, Machon P, Wolz M, S{\"{u}}rgers C, Beckmann D, Belzig W and Scheer E
  2018 {\em Nat. Commun.\/} {\bf 9} 5248 ISSN 2041-1723
  \urlprefix\url{http://www.nature.com/articles/s41467-018-07597-w}

\bibitem{Satchell2018}
Satchell N and Birge N~O 2018 {\em Phys. Rev. B\/} {\bf 97} 214509 ISSN
  2469-9950 \urlprefix\url{https://link.aps.org/doi/10.1103/PhysRevB.97.214509}

\bibitem{Satchell2019}
Satchell N, Loloee R and Birge N~O 2019 {\em Phys. Rev. B\/} {\bf 99} 174519
  ISSN 2469-9950
  \urlprefix\url{https://link.aps.org/doi/10.1103/PhysRevB.99.174519}

\bibitem{Eskilt2019}
Eskilt J~R, Amundsen M, Banerjee N and Linder J 2019 {\em Phys. Rev. B\/} {\bf
  100} 224519 ISSN 2469-9950
  \urlprefix\url{https://link.aps.org/doi/10.1103/PhysRevB.100.224519}

\bibitem{Bujnowski2019}
Bujnowski B, Biele R and Bergeret F~S 2019 {\em Phys. Rev. B\/} {\bf 100}
  224518 ISSN 2469-9950
  \urlprefix\url{https://link.aps.org/doi/10.1103/PhysRevB.100.224518}

\bibitem{Machida1978}
Machida K and Klemm R~A 1978 {\em Solid State Commun.\/} {\bf 27} 1061--1063
  ISSN 00381098
  \urlprefix\url{https://linkinghub.elsevier.com/retrieve/pii/0038109878911109}

\bibitem{Alidoust2014}
Alidoust M, Halterman K and Linder J 2014 {\em Phys. Rev. B\/} {\bf 89} 054508
  ISSN 1098-0121
  \urlprefix\url{https://link.aps.org/doi/10.1103/PhysRevB.89.054508}

\bibitem{Asano2014}
Asano Y, Fominov Y~V and Tanaka Y 2014 {\em Phys. Rev. B\/} {\bf 90} 094512
  ISSN 1098-0121
  \urlprefix\url{https://link.aps.org/doi/10.1103/PhysRevB.90.094512}

\bibitem{DiBernardo2015a}
{Di Bernardo} A, Salman Z, Wang X~L, Amado M, Egilmez M, Flokstra M~G, Suter A,
  Lee S~L, Zhao J~H, Prokscha T, Morenzoni E, Blamire M~G, Linder J and
  Robinson J~W~A 2015 {\em Phys. Rev. X\/} {\bf 5} 041021 ISSN 2160-3308
  \urlprefix\url{https://link.aps.org/doi/10.1103/PhysRevX.5.041021}

\bibitem{Rouco2019a}
Rouco M, Chakraborty S, Aikebaier F, Golovach V~N, Strambini E, Moodera J~S,
  Giazotto F, Heikkil{\"{a}} T~T and Bergeret F~S 2019 {\em Phys. Rev. B\/}
  {\bf 100} 184501 ISSN 2469-9950
  \urlprefix\url{https://doi.org/10.1103/PhysRevB.100.184501
  https://link.aps.org/doi/10.1103/PhysRevB.100.184501}

\bibitem{Meservey1994}
Meservey R and Tedrow P~M 1994 {\em Phys. Rep.\/} {\bf 238} 173--243
  \urlprefix\url{https://doi.org/10.1016/0370-1573(94)90105-8}

\bibitem{Tedrow1971}
Tedrow P~M and Meservey R 1971 {\em Phys. Rev. Lett.\/} {\bf 26} 192--195 ISSN
  00319007

\bibitem{Tedrow1973}
Tedrow P~M and Meservey R 1973 {\em Phys. Rev. B\/} {\bf 7} 318--326 ISSN
  01631829

\bibitem{Taylor1963}
Taylor B~N and Burstein E 1963 {\em Phys. Rev. Lett.\/} {\bf 10} 14--17 ISSN
  0031-9007 \urlprefix\url{https://link.aps.org/doi/10.1103/PhysRevLett.10.14}

\bibitem{Adkins1963}
Adkins C~J 1963 {\em Philos. Mag.\/} {\bf 8} 1051--1061 ISSN 0031-8086
  \urlprefix\url{http://www.tandfonline.com/doi/abs/10.1080/14786436308214463}

\bibitem{Adkins1964}
Adkins C~J 1964 {\em Rev. Mod. Phys.\/} {\bf 36} 211--213 ISSN 0034-6861
  \urlprefix\url{https://link.aps.org/doi/10.1103/RevModPhys.36.211}

\bibitem{VanHuffelen1993}
van Huffelen W~M, Klapwijk T~M, Heslinga D~R, de~Boer M~J and van~der Post N
  1993 {\em Phys. Rev. B\/} {\bf 47} 5170--5189 ISSN 0163-1829
  \urlprefix\url{https://link.aps.org/doi/10.1103/PhysRevB.47.5170}

\bibitem{Kuhlmann1994}
Kuhlmann M, Zimmermann U, Dikin D, Abens S, Keck K and Dmitriev V~M 1994 {\em
  Z. Phys. B\/} {\bf 96} 13--24 ISSN 0722-3277
  \urlprefix\url{http://link.springer.com/10.1007/BF01313010}

\bibitem{Zimmermann1995}
Zimmermann U, Abens S, Dikin D, Keck K and Dmitriev V~M 1995 {\em Z. Phys. B\/}
  {\bf 97} 59--66 ISSN 0722-3277
  \urlprefix\url{https://linkinghub.elsevier.com/retrieve/pii/0921453494921733
  http://link.springer.com/10.1007/BF01317588}

\bibitem{Rowell1964}
Rowell J~M 1964 {\em Rev. Mod. Phys.\/} {\bf 36} 199--200 ISSN 0034-6861
  \urlprefix\url{https://link.aps.org/doi/10.1103/RevModPhys.36.199}

\bibitem{Rowell1968}
Rowell J~M and Feldmann W~L 1968 {\em Phys. Rev.\/} {\bf 172} 393--401 ISSN
  0031-899X \urlprefix\url{https://link.aps.org/doi/10.1103/PhysRev.172.393}

\bibitem{Klapwijk1982}
Klapwijk T~M, Blonder G~E and Tinkham M 1982 {\em Physica\/} {\bf 109 {\&}
  110B} 1657--1664 \urlprefix\url{https://doi.org/10.1016/0378-4363(82)90189-9}

\bibitem{Octavio1983}
Octavio M, Tinkham M, Blonder G~E and Klapwijk T~M 1983 {\em Phys. Rev. B\/}
  {\bf 27} 6739--6746 ISSN 0163-1829
  \urlprefix\url{https://link.aps.org/doi/10.1103/PhysRevB.27.6739}

\bibitem{Flensberg1988}
Flensberg K, {Bindslev Hansen} J and Octavio M 1988 {\em Phys. Rev. B\/} {\bf
  38} 8707--8711 ISSN 0163-1829
  \urlprefix\url{https://link.aps.org/doi/10.1103/PhysRevB.38.8707}

\bibitem{Arnold1985}
Arnold G~B 1985 {\em J. Low Temp. Phys.\/} {\bf 59} 143--183 ISSN 0022-2291
  \urlprefix\url{http://link.springer.com/10.1007/BF00681510}

\bibitem{Arnold1987}
Arnold G~B 1987 {\em J. Low Temp. Phys.\/} {\bf 68} 1--27 ISSN 0022-2291
  \urlprefix\url{http://link.springer.com/10.1007/BF00682620}

\bibitem{Costa2021}
Costa A and Fabian J 2021 {\em Phys. Rev. B\/} {\bf 104} 174504
  \urlprefix\url{https://link.aps.org/doi/10.1103/PhysRevB.104.174504}

\bibitem{Gueron1996}
Gu{\'{e}}ron S, Pothier H, Birge N~O, Esteve D and Devoret M~H 1996 {\em Phys.
  Rev. Lett.\/} {\bf 77} 3025--3028 ISSN 0031-9007
  \urlprefix\url{https://link.aps.org/doi/10.1103/PhysRevLett.77.3025}

\bibitem{Pothier1997}
Pothier H, Gu{\'{e}}ron S, Birge N~O, Esteve D and Devoret M~H 1997 {\em Phys.
  Rev. Lett.\/} {\bf 79} 3490--3493 ISSN 0031-9007
  \urlprefix\url{https://link.aps.org/doi/10.1103/PhysRevLett.79.3490}

\bibitem{Kalcheim2015}
Kalcheim Y, Millo O, {Di Bernardo} A, Pal A and Robinson J~W~A 2015 {\em Phys.
  Rev. B\/} {\bf 92} 060501(R) ISSN 1098-0121
  \urlprefix\url{https://link.aps.org/doi/10.1103/PhysRevB.92.060501}

\bibitem{Ouassou2017a}
Ouassou J~A, Pal A, Blamire M, Eschrig M and Linder J 2017 {\em Sci. Rep.\/}
  {\bf 7} 1932 ISSN 2045-2322
  \urlprefix\url{http://dx.doi.org/10.1038/s41598-017-01330-1
  http://www.nature.com/articles/s41598-017-01330-1}

\bibitem{Alidoust2018}
Alidoust M and Halterman K 2018 {\em Phys. Rev. B\/} {\bf 97} 064517 ISSN
  2469-9950 \urlprefix\url{https://link.aps.org/doi/10.1103/PhysRevB.97.064517}

\bibitem{Halterman2018}
Halterman K and Alidoust M 2018 {\em Phys. Rev. B\/} {\bf 98} 134510
  \urlprefix\url{https://link.aps.org/doi/10.1103/PhysRevB.98.134510}

\bibitem{Alidoust2020}
Alidoust M and Halterman K 2020 {\em Phys. Rev. B\/} {\bf 102} 224504 ISSN
  2469-9950
  \urlprefix\url{https://link.aps.org/doi/10.1103/PhysRevB.102.224504}

\bibitem{Moodera1988}
Moodera J~S, Hao X, Gibson G~A and Meservey R 1988 {\em Phys. Rev. Lett.\/}
  {\bf 61} 637--640 ISSN 0031-9007
  \urlprefix\url{https://link.aps.org/doi/10.1103/PhysRevLett.61.637}

\bibitem{Strambini2017}
Strambini E, Golovach V~N, {De Simoni} G, Moodera J~S, Bergeret F~S and
  Giazotto F 2017 {\em Phys. Rev. Mater.\/} {\bf 1} 054402 ISSN 2475-9953
  \urlprefix\url{https://link.aps.org/doi/10.1103/PhysRevMaterials.1.054402}

\bibitem{DeSimoni2018}
{De Simoni} G, Strambini E, Moodera J~S, Bergeret F~S and Giazotto F 2018 {\em
  Nano Lett.\/} {\bf 18} 6369--6374 ISSN 1530-6984
  \urlprefix\url{https://pubs.acs.org/doi/10.1021/acs.nanolett.8b02723}

\bibitem{Lauter-Pasyuk2007}
Lauter-Pasyuk V 2007 {\em Collect. Soc. Fr. Neutron\/} {\bf 7} s221--s240

\bibitem{Lauter-Pasyuk1998}
Lauter-Pasyuk V, Lauter H~J, Aksenov V~L, Kornilov E~I, Petrenko A~V and
  Leiderer P 1998 {\em Physica B\/} {\bf 248} 166--170 ISSN 09214526
  \urlprefix\url{https://linkinghub.elsevier.com/retrieve/pii/S0921452698002269}

\bibitem{Lauter-Pasyuk1999}
Lauter-Pasyuk V, Lauter H~J, Lorenz M, Aksenov V~L and Leiderer P 1999 {\em
  Physica B\/} {\bf 267-268} 149--153 ISSN 09214526
  \urlprefix\url{https://linkinghub.elsevier.com/retrieve/pii/S0921452699000526}

\bibitem{Devizorova2019}
Devizorova Z, Mironov S~V, Mel'nikov A~S and Buzdin A 2019 {\em Phys. Rev. B\/}
  {\bf 99} 104519
  \urlprefix\url{https://link.aps.org/doi/10.1103/PhysRevB.99.104519}

\bibitem{Lauter-Pasyuk2009}
Lauter V, Ambaye H, Goyette R, {Hal Lee} W~T and Parizzi A 2009 {\em Physica
  B\/} {\bf 404} 2543--2546 ISSN 0921-4526
  \urlprefix\url{https://www.sciencedirect.com/science/article/pii/S092145260900369X}

\bibitem{Lauter-Pasyuk2000}
Lauter-Pasyuk V, Lauter H, Lorenz M, Petrenko A, Nikonov O, Aksenov V and
  Leiderer P 2000 {\em Physica B\/} {\bf 276-278} 776--777 ISSN 09214526
  \urlprefix\url{https://linkinghub.elsevier.com/retrieve/pii/S092145269901652X}

\bibitem{DeGennes1989}
De~Gennes P~G 1989 {\em {Superconductivity of Metals and Alloys}\/} (Addison
  Wesley, Redwood City)

\bibitem{Groth2014}
Groth C~W, Wimmer M, Akhmerov A~R and Waintal X 2014 {\em New J. Phys.\/} {\bf
  16} 063065 ISSN 1367-2630
  \urlprefix\url{http://stacks.iop.org/1367-2630/16/i=6/a=063065?key=crossref.4e4b2188754950ef7234ba8b61d069cb}

\bibitem{Csire2018}
Csire G, {\'{U}}jfalussy B and Annett J~F 2018 {\em Eur. Phys. J. B\/} {\bf 91}
  217 ISSN 1434-6028
  \urlprefix\url{http://link.springer.com/10.1140/epjb/e2018-90095-7}

\bibitem{Csire2018a}
Csire G, De{\'{a}}k A, Ny{\'{a}}ri B, Ebert H, Annett J~F and {\'{U}}jfalussy B
  2018 {\em Phys. Rev. B\/} {\bf 97} 024514 ISSN 2469-9950
  \urlprefix\url{https://link.aps.org/doi/10.1103/PhysRevB.97.024514}

\bibitem{Gmitra2013}
Gmitra M, Matos-Abiague A, Draxl C and Fabian J 2013 {\em Phys. Rev. Lett.\/}
  {\bf 111} 036603
  \urlprefix\url{http://link.aps.org/doi/10.1103/PhysRevLett.111.036603}

\end{thebibliography}

\end{document}